\documentclass[a4paper,11pt]{article}
\usepackage[toc,page]{appendix}
\usepackage{jheppub}
\usepackage{setspace}
\usepackage{amssymb}
\usepackage{amsmath,amsfonts}
\usepackage{graphicx}
\usepackage{mathrsfs}
\usepackage[vcentermath]{youngtab}
\usepackage{color}
\usepackage[normalem]{ulem}
\numberwithin{equation}{section}
\usepackage{bbm}
\newcommand{\nn}{\nonumber\\}

\newcommand{\be}{\begin{equation}}
\newcommand{\ee}{\end{equation}}
\newcommand{\bea}{\begin{eqnarray}}
\newcommand{\eea}{\end{eqnarray}}

\newcommand{\eq}{&=&}
\newcommand{\comm}[2]{\left[ #1 \, , \, #2 \right]}
\newcommand{\acomm}[2]{\left\{ #1 \, , \, #2 \right\}}
\newcommand{\ket}[1]{\left\lvert #1 \right\rangle}
\newcommand{\lb}{\left(}
\newcommand{\rb}{\right)}
\usepackage{multirow}
\newcommand{\half}{\frac{1}{2}}

\newcommand{\third}{\frac{1}{3}}
\newcommand{\sixth}{\frac{1}{6}}

\newcommand{\ml}{\mathcal{L}}

\renewcommand{\a}{\alpha}
\newcommand{\ad}{{\dot \alpha}}
\renewcommand{\b}{\beta}
\newcommand{\bd}{{\dot \beta}}
\renewcommand{\d}{\delta}
\newcommand{\g}{\gamma}
\newcommand{\gd}{{\dot \gamma}}
\newcommand{\dd}{{\dot \delta}}

\newcommand{\s}{{\,\,\, }}
\newcommand{\mfg}{\mathfrak{g}}

\newcommand{\bep}{\begin{picture}}
\newcommand{\eep}{\end{picture}}

\newcounter{YoungHeight}\newcounter{YoungWidth}

\newcounter{Mul1}\newcounter{Mul2}\newcounter{Mul3}\newcounter{Mul4}
\newcounter{A1}\newcounter{A2}
\newcounter{B3}
\newcounter{C3}\newcounter{C4}

\newcounter{T0}\newcounter{T1}

\newcounter{R0}

\newlength{\txtHShift}

\newlength{\txtWidth}

\newcommand{\Add}[3]{\setcounter{#1}{#2}\addtocounter{#1}{#3}}

\newcommand{\Length}[1]{#10}
\newcommand{\YoungScale}{}

\newcommand{\BlockA}[2]{{\YoungScale\bep(\Length{#1},\Length{#2}){\Add{A1}{#1}{1}\Add{A2}{#2}{1}}%
\multiput(0,0)(10,0){\value{A1}}{\line(0,1){\Length{#2}}}\multiput(0,0)(0,10){\value{A2}}{\line(1,0){\Length{#1}}}%
\setcounter{YoungHeight}{\Length{#2}}\setcounter{YoungWidth}{\Length{#1}}\eep}}

\newcommand{\YoungB}{\BlockA{2}{1}}

\newcommand{\YoungAA}{\BlockA{1}{2}}

\newcommand{\YoungBB}{\BlockA{2}{2}}

\newcommand{\YoungAAAA}{\BlockA{1}{4}}

\newcommand{\stln}{\setlength{\unitlength}{2.2ex}}
\newcommand{\fr}{\framebox(1,1){}}
\newcommand{\sdfr}{\framebox(1,1){\begin{picture}(1,1)
  \put(0,0){\line(1,1){1}}\put(0.25,0.25){$\bullet$}\end{picture}}}
  \newcommand{\sfr}{\framebox(1,1){\begin{picture}(1,1)
  \put(0,0){\line(1,1){1}}\end{picture}}}
  \newcommand{\dfr}{\framebox(1,1){\begin{picture}(1,1)
  \put(0.25,0.25){$\bullet$}\end{picture}}}

\newcommand{\oneonebox}
{\stln \lower2.6ex\hbox{
\begin{picture}(1.4,2.6)
\put(.3,.3){\fr}
\put(.3,1.3){\fr}
\end{picture}}}

\newcommand{\sonebox}
{\stln \lower2.6ex\hbox{
\begin{picture}(1.4,2.6)
\put(.3,1.1){\sfr}
\end{picture}}}

\newcommand{\donebox}
{\stln \lower2.6ex\hbox{
\begin{picture}(1.4,2.6)
\put(.3,1.1){\dfr}
\end{picture}}}

\newcommand{\onebox}
{\stln \lower2.6ex\hbox{
\begin{picture}(1.4,2.6)
\put(.3,1.1){\fr}
\end{picture}}}

\newcommand{\oneoneonebox}
{\stln \lower2.6ex\hbox{
\begin{picture}(1.4,3.6)
\put(.3,.3){\fr}
\put(.3,1.3){\fr}
\put(.3,2.3){\fr}
\end{picture}}}

\newcommand{\twotwobox}
{\stln \lower2.6ex\hbox{
\begin{picture}(2.6,2.6)
\put(.3,.3){\fr}
\put(.3,1.3){\fr}
\put(1.3,.3){\fr}
\put(1.3,1.3){\fr}
\end{picture}}}

\newcommand{\twooneonebox}
{\stln \lower2.6ex\hbox{
\begin{picture}(2.6,2.6)
\put(1.3,1.3){\fr}
\multiput(0.3,1.3)(0,-1){3}{\fr}
\end{picture}}}

\newcommand{\ttwooneonebox}
{\stln \lower2.6ex\hbox{
\begin{picture}(4.6,2.6)
\multiput(2.3,1.3)(1,0){2}{\fr}
\multiput(1.3,1.3)(0,-1){3}{\fr}
\multiput(0.3,1.3)(0,-1){3}{\fr}
\end{picture}}}

\newcommand{\stwooneonebox}
{\stln \lower2.6ex\hbox{
\begin{picture}(2.6,2.3)
\put(1.3,1.){\sfr}
\multiput(0.3,1.)(0,-1){1}{\sdfr}
\end{picture}}}

\newcommand{\dtwooneonebox}
{\stln \lower2.6ex\hbox{
\begin{picture}(2.6,2.3)
\put(1.3,1.){\fr}
\multiput(0.3,1.)(0,-1){1}{\dfr}
\end{picture}}}

\newcommand{\sttwooneonebox}
{\stln \lower2.6ex\hbox{
\begin{picture}(4.6,2.3)
\multiput(2.3,1.)(1,0){2}{\sfr}
\multiput(1.3,1.)(0,-1){1}{\sdfr}
\multiput(0.3,1.)(0,-1){1}{\sdfr}
\end{picture}}}

\newcommand{\marctwojtwojbox}
{\stln \lower2.6ex \hbox{
\begin{picture}(6.6, 4.1)
\multiput(.3, 1.3)(1,0) {2} {\fr}
\put(2.3,.3){\framebox(3,2){$\cdots$}}
\put(5.3, 1.3){\fr}
\multiput(.3,.3)(1,0){2}{\fr}
\put(5.3,.3){\fr}
\put(.3, 2.4){$\overbrace{~~~~~~~~~~~~~~~~~~}^{\text{two-row diagram}}$}
\end{picture}}}

\newcommand{\tmarctwojtwojbox}
{\stln \lower2.6ex \hbox{
\begin{picture}(12.6, 4.1)
\multiput(.3, 2.3)(1,0) {2} {\fr}
\multiput(.3, 1.3)(1,0) {2} {\fr}
\put(2.3,.3){\framebox(3,3){$\cdots$}}
\put(5.3, 1.3){\fr}
\put(5.3, 2.3){\fr}
\multiput(.3,.3)(1,0){2}{\fr}
\put(5.3,.3){\fr}
\multiput(6.3,2.3) (1, 0) {2} {\fr}
\put(8.3, 2.3) {\framebox(3,1){$\cdots$}}
\put(11.3, 2.3){\fr}
\end{picture}}}

\newcommand{\stmarctwojtwojbox}
{\stln \lower2.6ex \hbox{
\begin{picture}(12.6, 2.5)
\multiput(.3, 1.)(1,0) {2} {\sdfr}
\put(2.3,1.){\framebox(3,1){$\cdots$}}
\put(5.3, 1.){\sdfr}
\multiput(6.3,1.) (1, 0) {2} {\sfr}
\put(8.3, 1.) {\framebox(3,1){$\cdots$}}
\put(11.3, 1.){\sfr}
\end{picture}}}

\begin{document}

\begin{flushright}
 CERN-PH-TH/2013-296
\end{flushright}

\title{Deformed Twistors and Higher Spin Conformal (Super-)Algebras in Four Dimensions}

\author{Karan Govil$^a$  and  }
\author{Murat G\"unaydin$^{a,b}$}

\affiliation{$^a$ Institute for Gravitation and the Cosmos \\
 Physics Department ,
Pennsylvania State University\\
University Park, PA 16802, USA}
\affiliation{$^{a,b}$ Theory Division, Physics Department \\ CERN
CH-1211 Geneva,  Switzerland  }

\emailAdd{kzg126@psu.edu}
\emailAdd{murat@phys.psu.edu}
\abstract{Massless conformal scalar field in $d=4$ corresponds to the minimal unitary representation (minrep) of the conformal group $SU(2,2)$ which admits a one-parameter family of deformations that describe massless fields of arbitrary helicity. The minrep and its deformations  were obtained  by quantization of the nonlinear realization of $SU(2,2)$ as a quasiconformal group in arXiv:0908.3624.
We show that the generators of $SU(2,2)$ for these  unitary irreducible representations can be written as bilinears of \emph{deformed} twistorial oscillators which transform \emph{nonlinearly} under the Lorentz group and apply them to define and study higher spin algebras and superalgebras in $AdS_5$.
The higher spin (HS)  algebra of Fradkin-Vasiliev type in $AdS_5$ is simply  the enveloping algebra of $SU(2,2)$ quotiented by a two-sided ideal
(Joseph ideal) which annihilates the minrep. We show that the Joseph ideal vanishes identically for the quasiconformal realization  of the minrep and its enveloping algebra leads directly to the HS algebra in $AdS_5$.
Furthermore, the enveloping algebras of the deformations of the minrep define a one parameter family of HS algebras in $AdS_5$ for which certain $4d$ covariant  deformations of the Joseph ideal vanish identically. These results extend to superconformal algebras $SU(2,2|N)$ and  we find a one parameter family of HS superalgebras as enveloping algebras of the minimal unitary supermultiplet and its deformations.  Our results suggest the existence of a  family of (supersymmetric) HS theories in $AdS_5$ which are dual to free (super)conformal field theories (CFTs) or to  interacting but integrable (supersymmetric) CFTs  in $4d$. We also discuss the corresponding picture in HS algebras in $AdS_4$  where the corresponding $3d$ conformal group $Sp(4,\mathbb{R})$ admits  only two massless representations (minreps), namely the scalar and spinor singletons.

}

\maketitle

\section{Introduction}
Motivated  by the work of  physicists on spectrum generating symmetry groups in the 1960s  the concept of minimal unitary representations of noncompact Lie groups   was  introduced  by Joseph in \cite{MR0342049}. Minimal unitary representation of a  noncompact Lie group is defined over an Hilbert space of functions depending on the minimal number of variables possible. They have been studied extensively in the mathematics literature  \cite{MR644845,MR1103588,binegar1991unitarization,MR1159103,MR1372999,MR1278630,MR1327538,MR1737731,MR2020550,MR2020551,MR2020552,Gover:2009vc,Kazhdan:2001nx}.
A unified approach to the construction and study of minimal unitary representations of noncompact groups was developed after the discovery of novel  geometric quasiconformal realizations of noncompact groups in \cite{Gunaydin:2000xr}. Quasiconformal realizations exist for different  real forms of all noncompact groups as well as for their complex forms  \cite{Gunaydin:2000xr,Gunaydin:2005zz}\footnote{For the largest exceptional group  $E_{8(8)}$ the quasiconformal action  is the first and  only known geometric realization  of $E_{8(8)}$ and  leaves invariant a generalized light-cone with respect to a quartic distance function  in 57 dimensions  \cite{Gunaydin:2000xr}.}.

The quantization of  geometric quasiconformal action of a noncompact group leads directly to its minimal unitary representation as  was first shown  explicitly for the  split  exceptional group  $E_{8(8)}$ with the maximal compact subgroup $SO(16)$ \cite{Gunaydin:2001bt}. The minimal unitary representation  of  three dimensional U-duality group $E_{8(-24)}$ of the exceptional supergravity \cite{Gunaydin:1983rk} was similarly obtained  in \cite{Gunaydin:2004md}.
In \cite{Gunaydin:2006vz}  a unified formulation of the minimal unitary representations of  noncompact groups  based on the quasiconformal method was given and it was extended to the minimal representations of noncompact supergroups $G$ whose even subgroups are of the form $H\times SL(2,\mathbb{R})$ with $H$ compact\footnote{We shall be mainly working at the level of  Lie algebras and Lie superalgebras and be cavalier about using the same symbol to denote a (super)group and its Lie (super)algebra.} . These supergroups include  $G(3)$ with even subgroup $G_2\times SL(2,\mathbb{R})$, $F(4)$ with even subgroup $Spin(7)\times SL(2,\mathbb{R}) $, $D\left(2,1;\sigma\right)$ with even subgroup $ SU(2)\times SU(2) \times SU(1,1)$  and $OSp\left(N|2,\mathbb{R}\right)$. These results were further generalized to supergroups of the form $SU(n,m|p+q)$ and $OSp(2N^*|2M)$ in \cite{Fernando:2009fq,Fernando:2010dp,Fernando:2010ia} and applied to conformal superalgebras in 4 and 6 dimensions. In particular, the construction of the minreps of  $5d$ anti-de Sitter  or $4d$ conformal group $SU(2,2)$ and  corresponding supergroups $SU(2,2|N)$ was given in  \cite{Fernando:2009fq}. One finds that the minimal unitary representation of the group $SU(2,2)$ obtained  by quantization of its quasiconformal realization is isomorphic to the scalar doubleton representation that describes a massless scalar field in four dimensions. Furthermore  the minrep of $SU(2,2)$ admits a one parameter family ($\zeta$) of deformations that can be identified with helicity, which can be continuous. For a positive (negative)  integer value of the deformation parameter  $\zeta$,  the resulting unitary irreducible representation of  $SU(2,2)$ corresponds to a $4d$ massless conformal field transforming in $\left(0 \,,\, \frac{\zeta}{2} \right)$ $\lb \lb -\frac{\zeta}{2},0 \rb\rb$ representation of the Lorentz subgroup, $SL(2,\mathbb{C})$. These deformed minimal representations for integer values of $\zeta$  turn out to be isomorphic to the doubletons  of $SU(2,2)$ \cite{Gunaydin:1984fk,Gunaydin:1998sw,Gunaydin:1998jc}.

The minimal unitary supermultiplet of $SU(2,2|N)$ is the CPT self-conjugate (scalar) doubleton supermultiplet, and for $PSU(2,2|4)$ it is simply the four dimensional $N=4$  Yang-Mills supermultiplet. One finds that there exists a one-parameter family of deformations of the minimal unitary supermultiplet of $SU(2,2|N)$ .  The minimal unitary supermultiplet of $SU(2,2\,|\,N)$ and its deformations with integer $\zeta$  are isomorphic to  the unitary doubleton supermultiplets studied in \cite{Gunaydin:1984fk,Gunaydin:1998sw,Gunaydin:1998jc}.

The minrep of $7d$ $AdS$ or $6d$ conformal group $SO(6,2)=SO^*(8)$ and its deformations were studied in \cite{Fernando:2010dp}. One finds that the minrep admits deformations labelled by the eigenvalues of the Casimir of an $SU(2)_T$ subgroup of the little group, $SO(4)$, of massless particles in six dimensions. These deformed minreps labeled by spin $t$ of $SU(2)_T$ are positive energy unitary irreducible representations of $SO^*(8)$ that describe massless conformal fields in six dimensions.
Quasiconformal construction of the minimal unitary supermultiplet of $OSp(8^*|2N)$  and its deformations were given in \cite{Fernando:2010dp,Fernando:2010ia}. The minimal unitary supermultiplet of $OSp(8^*|4)$ is the massless conformal $(2,0)$ supermultiplet whose interacting theory  is believed to be dual to M-theory on $AdS_7 \times S^4$ . It is isomorphic to the scalar doubleton supermultiplet of  $OSp(8^*|4)$ first constructed in \cite{Gunaydin:1984wc}.

For symplectic groups $Sp(2N,\mathbb{R})$ the construction of the minimal unitary representation using the quasiconformal approach and the covariant twistorial oscillator method coincide \cite{Gunaydin:2006vz}. This is due to the fact that the quartic invariant operator that enters the quasiconformal construction vanishes  for symplectic groups and hence the resulting generators involve only bilinears of oscillators. Therefore the minreps of $Sp(4,\mathbb{R})$ are simply the scalar and spinor singletons that were called the remarkable representations of anti-de Sitter group by Dirac \cite{dirac1963remarkable}. Unitary supermultiplets of general spacetime superalgebras were first constructed using the oscillator method developed in \cite{Gunaydin:1981yq,Bars:1982ep}. The singleton supermultiplets of  $AdS_4$ superalgebras , in particular those of  $N=8$ superalgebra $OSp(8|4)$ were first constructed, using the oscillator method,  in \cite{Gunaydin:1983cc,Gunaydin:1983yj}. Scalar and spinor singletons as representation of the $N=1$ , $AdS_4$ super algebra $OSp(1/4,\mathbb{R})$ were studied by Fronsdal \cite{Fronsdal:1981gq}
who called it a Dirac supermultiplet.
The oscillator construction  of the general unitary representations of $OSp(N|4,\mathbb{R})$  was further developed in \cite{Gunaydin:1984wc,Gunaydin:1988kz}.
The singleton supermultiplets of $OSp(N/4,\mathbb{R})$ were also studied in \cite{Nicolai:1984gb}.

The Kaluza-Klein spectrum of IIB supergravity over the $AdS_5 \times S^5$ space was first obtained via the twistorial oscillator method by  tensoring of the CPT self-conjugate doubleton supermultiplet of $SU(2,2\,|\,4)$ with itself repeatedly and restricting to the CPT self-conjugate short supermultiplets  \cite{Gunaydin:1984fk}.  Authors of  \cite{Gunaydin:1984fk} also pointed out that the  CPT self-conjugate doubleton supermultiplet $SU(2,2\,|\,4)$ does not have a Poincar\'{e} limit in five dimensions and its field theory  lives on the boundary of $AdS_5$ on which $SU(2,2)$ acts as a conformal group and that the unique candidate for this theory is the  four dimensional $N=4$ super Yang-Mills theory that is  conformally invariant. Similarly the Kaluza-Klein spectra of the compactifications of 11 dimensional supergravity over $AdS_4\times S^7$ and $AdS_7 \times S^4$ were obtained by tensoring  of singleton supermultiplet of   $OSp(8\,|\,4,\mathbb{R})$  \cite{Gunaydin:1985tc} and of scalar doubleton supermultiplet of  $OSp(8^*\,|\,4)$ \cite{Gunaydin:1984wc}, respectively. The authors of   \cite{Gunaydin:1985tc} and \cite{Gunaydin:1984wc} also pointed out that the field  theories of the singleton and scalar doubleton supermultiplets live on the boundaries of $AdS_4$ and $AdS_7$ as conformally invariant field theories, respectively\footnote{That the Poincare limit of singletons of $Sp(4,\mathbb{R})$ is singular and their local field  theories must be formulated on the boundary of $AdS_4$ as conformally invariant field theirs was first pointed out by Fronsdal and collaborators. See  \cite{Fronsdal:1981gq} and the references therein .}.  As such these works represent some of the earliest work on $AdS/CFT$ dualities within the framework of Kaluza-Klein supergravity theories. Their extension to the superstring and M-theory arena \cite{Maldacena:1997re,Witten:1998qj,Gubser:2002tv} started the modern era of $AdS/CFT$ research.  The fact that the scalar doubleton supermultiplets of $SU(2,2|4)$ and $OSp(8^*\,|\,4)$ and the singleton  supermultiplet of $OSp(8\,|\,4,\mathbb{R})$ turn out to be  the minimal unitary supermultiplets show that they are  very special from a mathematical point of view as well.

Tensor product  of the two singleton representations of the $AdS_4$ group $Sp(4,\mathbb{R})$ decomposes into infinitely many massless spin representations in $AdS_4$ as was shown in \cite{Flato:1978qz}. These higher spin theories were studied by Fronsdal and collaborators \cite{flato1981quantum,Fronsdal:1981gq,Flato:1980we,Angelopoulos:1980wg}. In the eighties Fradkin and Vasiliev initiated the study of higher spin theories involving fields of all spins $0 \leq s < \infty $ \cite{Fradkin:1986qy,Konshtein:1988yg}.  A great deal of work was done on higher spin theories since then and for comprehensive reviews on higher spin  theories we refer to \cite{Vasiliev:1995dn,Vasiliev:1999ba,Bekaert:2005vh,Iazeolla:2008bp,Sagnotti:2011qp} and references therein. The work on higher spin theories has  intensified in the last decade since the conjectured duality between Vasiliev's higher spin gauge theory in $AdS_4$ and $O(N)$ vector models in \cite{Klebanov:2002ja,Sezgin:2003pt}. The three point functions of higher spin currents were computed directly and matched with those of free and critical $O(N)$ vector models in \cite{Giombi:2009wh,Giombi:2010vg}. Substantial work has also been done in higher spin holography in the last few years and for references  we refer to the review \cite{Giombi:2012ms}.

In the early days of higher spin theories
it was  pointed out in \cite{Gunaydin:1989um} that the Fradkin-Vasiliev higher spin algebra in $AdS_4$ \cite{Fradkin:1986qy} corresponds simply to the infinite dimensional Lie algebra defined by the enveloping algebra of the singletonic realization of $Sp(4,\mathbb{R})$  and that this can be extended to construction of HS algebras  in higher dimensions.
The oscillator construction of the singleton representations \cite{Gunaydin:1983yj,Gunaydin:1983cc} of $AdS_4$ superalgebras were used in the  study of higher spin (super)algebras in \cite{Konshtein:1988yg}, where the admissibility condition was formulated. Conformal higher spin superalgebras were studied shortly thereafter in \cite{Fradkin:1989yd}.
  Again in \cite{Gunaydin:1989um} it was pointed out that  the supersymmetric extensions of the higher spin algebras in $AdS_4$, $AdS_5$ and $AdS_7$ could be similarly constructed as enveloping algebras of the singletonic or doubletonic realizations of the super algebras $OSp(N/4,\mathbb{R})$, $SU(2,2|N)$  and $OSp(8^*|2N)$.
  Higher spin algebras and superalgebras in $AdS_5$ and $AdS_7$ were studied along these lines in \cite{Sezgin:2001zs,Sezgin:2001ij,Sezgin:2001yf,Sezgin:2002rt} using the doubletonic  realizations of underlying algebras and superalgebras given in \cite{Gunaydin:1984wc,Gunaydin:1998jc,Gunaydin:1998sw,Fernando:2001ak}.  Higher spin superalgebras in dimensions $d > 3$ were also studied by Vasiliev in \cite{Vasiliev:2004cm}. However, they do not have the standard finite dimensional $AdS$ superalgebras as subalgebras  except for the case of $AdS_4$. The relation between higher spin algebras and cubic interactions for simple mixed-symmetry fields in $AdS$ space times using Vasiliev's approach was studied in \cite{Boulanger:2011se}.

Mikhailov  showed the connection between  $AdS_5/Conf_4$ higher spin algebra and the algebra of \emph{conformal Killing vectors and Killing tensors} in $d=4$ and their relation to to  higher symmetries of the Laplacian \cite{Mikhailov:2002bp} . The connection between conformal Killing vectors and tensors and higher symmetries of the Laplacian was put on a firm mathematical foundation by Eastwood \cite{Eastwood:2002su} who gave a realization of the $AdS_{(d+1)}/CFT_d$ higher spin algebra   as an explicit quotient of the universal enveloping algebra $\mathscr{U}(\mfg)$, ($\mfg = \mathfrak{so}(d,2))$, by a two-sided ideal $\mathscr{J}(\mfg)$ . This ideal $\mathscr{J}(\mfg)$ was identified as the annihilator of the scalar singleton module or the minimal representation and is known as the Joseph ideal in the mathematics  literature \cite{joseph1976minimal}

 This result  agrees with the proposal of \cite{Gunaydin:1989um} for $AdS_4/CFT_3$ higher spin algebras since singletons are simply the minreps of  $SO(3,2)$ and in its  singletonic twistorial oscillator realization the Joseph ideal vanishes identically as discussed in section \ref{singletons}. The  covariant  twistorial oscillators  have been used extensively in the formulation and study of higher spin $AdS_4$ algebras  since the early work of Fradkin and Vasiliev.  For the doubletonic realization of $SO(4,2)$ and  $SO(6,2)$ in terms of covariant twistorial oscillators the two sided Joseph ideal does not vanish identically as operators and must be quotiented out.  However, as will be shown explicitly in section \ref{josephqc4d},  the Joseph ideal vanishes identically as operators for the minimal unitary realization of $SU(2,2)$ obtained via quasiconformal approach. The  same result holds true for the $AdS_7/Conf_6$ algebras \cite{Govil:2014uwa}.

  One of the key results of this paper is to show that the basic objects for the construction of irreducible higher spin $AdS_5/Conf_4$ algebras  are not the covariant twistorial oscillators but rather the deformed twistorial oscillators that transform nonlinearly under the Lorentz group $SL(2,\mathbb{C})$. The minimal unitary representation of $SO(4,2)$ and its deformations obtained by quasiconformal methods \cite{Fernando:2009fq} can be written  as bilinears of these deformed twistors.  One parameter family of  higher spin $AdS_5/Conf_4$ algebras and superalgebras can thus be realized as enveloping algebras  involving products of  bilinears of  these deformed twistors\footnote{ Our results for deformed twistorial oscillators extend to higher spin superalgebras in $d=6$ and their deformations\cite{Govil:2014uwa}.}.
   We shall also review the $AdS_4/CFT_3$ algebras and their supersymmetric extensions so as to highlight the differences with the higher dimensional algebras. $AdS_4$ group $SO(3,2)$ is isomorphic to the symplectic group $Sp(4,\mathbb{R})$  and as was shown in \cite{Gunaydin:2006vz} the quasiconformal realization of symplectic groups reduce to realization in terms of bilinears of covariant oscillators.

The plan of the paper is as follows: In section \ref{sect-3dalgebra} we review the covariant twistorial oscillator (singleton) construction of the conformal group in three dimensions $SO(3,2)\sim Sp(4,\mathbb{R})$ and its superextension $OSp(N|4,\mathbb{R})$. Then we review the covariant twistorial oscillator (doubleton) construction for the four dimensional conformal group $SO(4,2) \sim SU(2,2)$ in section \ref{4d-doubleton-section}. In section \ref{sect4dqc}, we present the minimal unitary representation of $SU(2,2)$ obtained by the quasiconformal approach \cite{Fernando:2009fq} in terms of certain deformed twistorial oscillators that transform nonlinearly under the Lorentz group.  We then define a one-parameter family of these deformed twistors, which we call \emph{helicity deformed}  twistorial oscillators and  express the generators of  a one parameter family of deformations of the minrep given in \cite{Fernando:2009fq} as bilinears of the helicity deformed twistors.  They describe massless  conformal fields of arbitrary helicity which can be continuous. In section \ref{section-superconformal}, we use the deformed twistors to realize the superconformal algebra $PSU(2,2|4)$ and its deformations in the quasiconformal framework.  In section \ref{sect-joseph}, we review the Eastwood's formula for the  generator $\mathscr{J}$ of the annihilator of the minrep (Joseph ideal) and show by explicit calculations that it vanishes identically for the singletons of $SO(3,2)$ and the minrep of $SU(2,2)$ obtained by quasiconformal methods. We then present the generator  $\mathscr{J}$ of the Joseph ideal in $4d$ covariant indices and use them to define the deformations $\mathscr{J}_\zeta$ that are the annihilators of the deformations of the minrep. In section \ref{hssect}, we use the fact that annihilators vanish identically to identify the $AdS_5/Conf_4$ higher spin algebra (as defined by Eastwood \cite{Eastwood:2002su}) and define its deformations as the enveloping algebras of the deformations of the minrep within the quasiconformal framework. In  section 4.4 we discuss the extension of these results to higher spin superalgebras. Finally in section \ref{section-discussion} we discuss  the implications of our results for higher spin theories of massless fields in $AdS_5$ and their conformal duals in $4d$.

\section{$3d$ conformal algebra $SO(3,2)\sim Sp(4,\mathbb{R})$ and its minimal unitary realization}
\label{sect-3dalgebra}

In this section we shall review the twistorial oscillator construction of the unitary representations of the conformal groups $SO(3,2)$ in $d=3$ dimensions that correspond to conformally massless fields in $d=3$ following \cite{Gunaydin:1985tc,Chiodaroli:2011pp}. These representations turn out to be the minimal unitary representations and are also called the singleton (scalar and spinor singleton) representations of Dirac \cite{dirac1963remarkable}. The quasiconformal and covariant oscillator rconstructions of symplectic groups $Sp(2N,\mathbb{R})$  coincide\cite{Gunaydin:2006vz} and thus we will only review the oscillator construction of $Sp(4,\mathbb{R})$ following \cite{Gunaydin:1985tc,Chiodaroli:2011pp}.
 \subsection{Twistorial oscillator construction of $SO(3,2)$}
 \label{so32twist}

The covering group of the three (four) dimensional conformal (anti-de Sitter) group $SO(3,2)$  is isomorphic to the noncompact symplectic group $Sp(4,\mathbb{R})$ with the maximal compact subgroup $U(2)$.
Commutation relations of its generators  can be written as
\be
\comm{M_{AB}}{M_{CD}} = i (\eta_{BC}M_{AD}-\eta_{AC}M_{BD}-\eta_{BD}M_{AC}+\eta_{AD}M_{BC}) \label{so32comm}
\ee
where $\eta_{AB} = \text{diag}(-,+,+,+,-)$ and $A,B = 0,1, \ldots, 4$.
Spinor representation of $SO(3,2)$ can be realized in terms of  four-dimensional gamma matrices $\gamma_\mu$ that satisfy \[ \acomm{\gamma_\mu}{\gamma_\nu} = -2\eta_{\mu\nu}\] where $\eta_{\mu\nu} = \text{diag}(-,+,+)$ and $\mu,\nu = 0, 1, \ldots, 3$ and $\gamma_5 = \gamma_0 \gamma_1 \gamma_2\gamma_3$ as follows:

\be
\Sigma_{\mu\nu} := -\frac{i}{4}\comm{\gamma_\mu}{\gamma_\nu}, \qquad\qquad \Sigma_{\mu 4} := -\frac{1}{2}\gamma_\mu
\ee
We adopt the following conventions for gamma matrices in four dimensions:
\be
\gamma_0 = \begin{pmatrix}
    \mathbbm{1}_2 & 0  \\
    0 & -\mathbbm{1}_2
\end{pmatrix},
\qquad \qquad
\gamma_m = \begin{pmatrix}
   0 &  -\sigma_m   \\
     \sigma_m & 0
\end{pmatrix},
\qquad\qquad
\gamma_5 = i\begin{pmatrix}
    \mathbbm{1}_2 & 0  \\
    0 & \mathbbm{1}_2
\end{pmatrix}
\ee
where $\sigma_m$ ($m=1,2,3$ are Pauli matrices.
 Consider now a pair of bosonic oscillators $a_i,a_i^\dagger$ ($\, i=1,2$) that satisfy
\be
\comm{a_i}{a_j^\dagger} = \delta_{ij}.
\ee
 and define a twistorial (Majorana) spinor $\Psi$ and its Dirac conjugate in terms of these oscillators
 $ \overline{\Psi} = \Psi^\dagger \gamma_0 $
\be
\Psi = \begin{pmatrix}
a_1\\ -ia_2 \\ ia_2^\dagger \\ -a_1^\dagger
\end{pmatrix},
\qquad
\overline{\Psi} = \begin{pmatrix}
a_1^\dagger \,\, ia_2^\dagger \,\,  ia_2 \,\,  a_1
\end{pmatrix}
\ee
Then the bilinears $M_{AB} =2 \overline{\Psi} \Sigma_{AB} \Psi$ satisfy the commutation relations (\ref{so32comm}) of $SO(3,2)$ Lie  algebra.

The Fock space of these oscillators decompose into two ireducible unitary representations of $Sp(4,\mathbb{R})$ that are simply the two remarkable representations of Dirac \cite{dirac1963remarkable} which were called $Di$ and $Rac$ in \cite{flato1981quantum}. These representations do not have a Poincar\'e limit in $4d$ and their field theories live on the boundary of $AdS_4$ which can be identified with the conformal compactification of three dimensional Minkowski space \cite{Angelopoulos:1980wg}.

\subsection{$SO(3,2)$ algebra in conformal three-grading and $3d$ covariant twistors}
The conformal algebra in $d$ dimensions can be given a three graded decomposition with respect to the noncompact dilatation generator $\Delta$ as follows:
\be
\mathfrak{so}(d,2) = K_\mu \oplus \lb M_{\mu\nu} \oplus \Delta \rb \oplus P_\mu
\label{3dconfgrading}
\ee
We shall call this conformal 3-grading. The commutation relations of the algebra in this basis are given as follows:
\begin{equation}
\begin{split}
\comm{M_{\mu\nu}}{M_{\rho\tau}}
&= i \left(
     \eta_{\nu\rho}M_{\mu\tau}
     - \eta_{\mu\rho} M_{\nu\tau}
     - \eta_{\nu\tau} M_{\mu\rho}
     + \eta_{\mu\tau} M_{\nu\rho}
    \right)
\\
\comm{P_\mu}{M_{\nu\rho}}
&= i \left( \eta_{\mu\nu} \, P_\rho
            - \eta_{\mu\rho} \, P_\nu
     \right)
\\
\comm{K_\mu}{M_{\nu\rho}}
&= i \left( \eta_{\mu\nu} \, K_\rho
            - \eta_{\mu\rho} \, K_\nu
     \right)
\\
\comm{\Delta}{M_{\mu\nu}}
&= \comm{P_\mu}{P_\nu}
 = \comm{K_\mu}{K_\nu}
 = 0
\\
\comm{\Delta}{P_\mu}
&= + i \, P_\mu
\qquad \qquad
\comm{\Delta}{K_\mu}
 = - i \, K_\mu
\\
\comm{P_\mu}{K_\nu}
&= 2 i \left( \eta_{\mu\nu} \, \Delta + M_{\mu\nu} \right)
\end{split}
\end{equation}
where $M_{\mu\nu}$ ($\mu,\nu=0,1,...,(d-1)$) are the Lorentz groups generators. $P_\mu$ and $K_\mu$ are the generators of translations and special conformal transformations.

In $d=3$ dimensions the Greek indices $\mu,\nu,...$ run over $0,1,2$ and dilatation generator is simply
\be
D=  -M_{34}
\ee
and translations $P_\mu$ and special conformal transformations
$K_\mu$ are given by:
\bea
P_\mu\eq M_{\mu 4}+ M_{\mu 3} \\
K_\mu \eq M_{\mu 4}-M_{\mu 3}
\label{so42-3grading2}
\eea

In order to make connection with higher spin (super-)algebras it is best to write the algebra in $SO(2,1)$ covariant spinorial oscillators. Let us now introduce linear combinations of $a_i,a_i^\dagger$ which we shall call $3d$ twistors \footnote{Note that the $3d$ twistor variables defined in \cite{Chiodaroli:2011pp} look slightly different from the ones defined here because the oscillators $a_i,a^i$ are linear combinations of the ones used in \cite{Chiodaroli:2011pp}.}:
\begin{align}
\kappa_1 &= \frac{i}{2} \lb a_1+a_2^\dagger+a_1^\dagger+a_2 \rb, \qquad \mu^1 = \frac{i}{2} \lb a_1-a_2^\dagger-a_1^\dagger+a_2 \rb \\
\kappa_2 &= \frac{1}{2} \lb a_1+a_2^\dagger-a_1^\dagger-a_2 \rb, \qquad \mu^2 = \frac{1}{2} \lb a_1-a_2^\dagger+a_1^\dagger-a_2 \rb
\end{align}
They satisfy the following commutation relations:
\be
\comm{\kappa_\a}{\mu^\b} = \d_\a^\b
\ee
Using these we can write (spinor conventions for $SO(2,1)$ are given in appendix \ref{so21}):
\bea
P_{\a\b} = \lb \sigma^\mu P_\mu \rb_{\a\b} \eq -\kappa_\a \kappa_\b \\
K^{\a\b} = \lb\bar{\sigma}^\mu K_\mu \rb^{\a\b} \eq -\mu^\a \mu^\b
\eea

Similarly we can define the Lorentz generators
\bea
M_\a^{\s \b} \eq i \lb \sigma^\mu\ \bar{\sigma}^\nu \rb_\a^{\s \b} M_{\mu\nu}  \\
\eq \kappa_\a \mu^\b - \half \d_\a^\b \kappa_\g \mu^\g
\eea
and the dilatation generator
\be
\Delta = -\frac{i}{4} \lb \kappa_\a \mu^\a + \mu^\a \kappa_\a \rb
\ee

In this basis the conformal algebra becomes:
\bea
\comm{M_\a^{\s\b}}{M_\g^{\s\d}} \eq \d_\a^\d M_\g^{\s\b} - \d_\g^\b M_\a^{\s\d} \\
\comm{P_{\a\b}}{M_\g^{\s\d}} \eq 2\d_{(\a}^\d P_{\b)\g} - \d_\g^\d P_{\a\b} \\
\comm{K^{\a\b}}{M_\g^{\s\d}} \eq -2\d^{(\a}_\g K^{\b)\d} + \d_\g^\d K^{\a\b} \\
\comm{P_{\a\b}}{K^{\g\d}} \eq 4 \d_{(\a}^{(\g} M_{\b)}^{\s \d)} + 4i \d_{(\a}^{(\g} \d_{\b)}^{\d)} \Delta
\eea
\be
\comm{\Delta}{K^{\a \b}} = -iK^{\a \b}, \quad \comm{\Delta}{M_\a^{\s \b}}=0, \quad \comm{\Delta}{P_{\a \b}} = iP_{\a \b}
\ee

The  conformal group $Sp(4,\mathbb{R})$ in three dimensions admits extensions to supergroups $OSp(N|4,\mathbb{R})$ with even subgroups $Sp(4,\mathbb{R})\times O(N)$. We review the minimal unitary realization of  $OSp(N|4,\mathbb{R})$ in Appendix \ref{OSp_N_4}.

\section{Conformal and superconformal algebras in four dimensions}
\label{sect-4dalgebra}

In this section, we present two different realizations of the conformal algebra and its supersymmetric extensions in $d=4$. We start by reviewing the doubleton oscillator realization \cite{Gunaydin:1984fk,Gunaydin:1998jc,Gunaydin:1998sw} and its reformulation  in terms of Lorentz covariant twistorial oscillators \cite{Gunaydin:1998sw,Chiodaroli:2011pp}. We then present a novel formulation of the quasiconformal realization of the minimal unitary representation and its deformations first studied in \cite{Fernando:2009fq} in terms of deformed twistorial oscillators.

\subsection{Covariant twistorial oscillator construction of the doubletons of  $SO(4,2)$}
\label{4d-doubleton-section}
The covering group  of the conformal group $SO(4,2)$ in four dimensions is   $SU(2,2)$. Denoting its generators as $M_{AB}$ the commutation relations in the canonical basis are
\be
\comm{M_{AB}}{M_{CD}} = i (\eta_{BC}M_{AD}-\eta_{AC}M_{BD}-\eta_{BD}M_{AC}+\eta_{AD}M_{BC}) \nn
\ee
where $\eta_{AB} = \text{diag}(-,+,+,+,+,-)$ and $A,B = 0, \ldots,
5$.
The spinor representation of $SO(4,2)$ can be realized in terms in four-dimensional
gamma matrices $\gamma_\mu$ that satisfy \be \acomm{\gamma_\mu}{\gamma_\nu}
= -2\eta_{\mu\nu}\ee where $\eta_{\mu\nu} = \text{diag}(-,+,+,+)$ ( $\mu,\nu
= 0, \ldots, 3$) as follows:

\be
\Sigma_{\mu\nu} := -\frac{i}{4}\comm{\gamma_\mu}{\gamma_\nu}, \qquad \Sigma_{\mu 4
} := -\frac{i}{2}\gamma_\mu\gamma_5, \qquad \Sigma_{\mu 5} := -\frac{1}{2}\gamma_\mu,
\qquad \Sigma_{45} := -\frac{1}{2}\gamma_5
\ee

Consider now two pairs of bosonic oscillators $a_i,a_j^\dagger$ ($i,j=1,2$)
and $b_r,b_s^\dagger$ ($r,s=1,2$) that satisfy
\be
\comm{a_i}{a_j^\dagger} = \delta_{ij}, \qquad \comm{b_r}{b_s^\dagger} = \delta_{rs}
\ee
 We form a twistorial  Dirac spinor $\Psi$ and its  conjugate $\overline{\Psi}
= \Psi^\dagger \gamma_0$ in terms of these oscillators:
\be
\Psi = \begin{pmatrix}
a_1\\ a_2 \\ -b_1^\dagger \\ -b_2^\dagger
\end{pmatrix},
\qquad
\overline{\Psi} = \begin{pmatrix}
a_1^\dagger \,\, a_2^\dagger \,\,  b_1 \,\,  b_2
\end{pmatrix}
\ee
Then the bilinears $M_{AB} =\overline{\Psi} \Sigma_{AB} \Psi$ ($A,B =
0, \ldots, 5$) generate  the Lie algebra of $SO(4,2)$:
\be
\comm{\,\overline{\Psi} \Sigma_{AB} \Psi}{\overline{\Psi} \Sigma_{CD}
\Psi} = \overline{\Psi} \comm{\Sigma_{AB}}{\Sigma_{CD}} \Psi
\ee
which was called the doubleton realization \cite{Gunaydin:1984fk,Gunaydin:1998jc,Gunaydin:1998sw}\footnote{The term doubleton refers to the fact that we are using oscillators that decompose into two irreps under the action of the maximal compact subgroup. For $SU(2,2)$ that is the minimal set required. For symplectic groups the minimal set consists of oscillators that form a single irrep of their maximal compact subgroups.}.

The Lie algebra of $SU(2,2)$ can be given a  three-grading with respect to the algebra of its maximal compact subgroup $SU(2)_L\times SU(2)_R\times U(1)$
\be
\mathfrak{su}(2,2) = L_{ir} \oplus ( L^i_j + R^i_j + E ) \oplus L^{ir}
\ee
which is referred to as  the compact three-grading.
 For the doubleton realizations  one has
\begin{align}
L^i_j &=  a^i a_j -\frac{1}{2} \delta^i_j (a^ka_k ), \qquad R^r_s  = b^r b_s -\frac{1}{2} \delta^r_s (b^t b_t )  \\
L_{ir}  &= a_i b_r, \quad E =\frac{1}{2} ( a^i a_i + b_r b^r ), \quad L^{ir}  =  a^i b^r
\end{align}
where  the creation operators are denoted with upper indices, i.e $a_i^\dagger=a^i$. Under the
$SU(2)_L\times SU(2)_R$ subgroup of $SU(2,2)$ generated by the bilinears $L^i_j$ and $R^r_s$ oscillators $a_i (a_i^\dagger)$
and $b_r(b_r^\dagger)$ transform in the $(1/2,0)$ and $(0,1/2)$ representation.
In contrast to the  situation in three dimensions, the Fock space of these bosonic oscillators decomposes into an  {\it infinite}  set
of positive energy unitary irreducible representations (UIRs), called doubletons  of $SU(2,2)$. These UIRs are uniquely
determined by  a subset of states with the lowest eigenvalue (energy) of the $U(1)$ generator and transforming irreducibly under  the $SU(2)_L\times
SU(2)_R$ subgroup.
The possible lowest
energy irreps of $SU(2)_L\times SU(2)_R$ for  positive energy UIRs of $SU(2,2)$ are of the form
\bea
a_{i_1}^\dagger a_{i_2}^\dagger \cdots a_{i_n}^\dagger \ket{0} &  \Leftrightarrow
  & (j_L,j_R)=\lb \frac{n}{2},0 \rb \qquad \text{E}=1+n/2 \\
b_{r_1}^\dagger b_{r_2}^\dagger \cdots b_{r_m}^\dagger \ket{0} &  \Leftrightarrow
  & (j_L,j_R)=\lb 0,\frac{m}{2} \rb \qquad \text{E}=1+m/2
\eea
It is worth mentioning that the doubleton representations are massless in four dimensions and their tensor products decompose into an infinite set of massless spin representations in $AdS_5$ \cite{Gunaydin:1984fk,Gunaydin:1998jc,Gunaydin:1998sw}. The tensoring procedure is straightforward in oscillator construction and it just corresponds to taking two copies ( colors)  of oscillators $a_i(\xi), a^i(\xi),b_i(\eta), b^i(\eta)$ where $\xi, \eta =1,2$. The resulting representations are multiplicity free\footnote{The explicit formulas for the tensor product decompositions of two irreducible doubleton representations  were given in \cite{Heidenreich:1980xi}.}. Tensoring more that two copies of doubleton irreps  decomposes into an infinite set of  massive representations in $AdS_5$ which are also multiplicity free\cite{Gunaydin:1984fk,Gunaydin:1998jc,Gunaydin:1998sw}.

To relate the oscillators transforming covariantly under the maximal compact subgroup $SO(4)\times U(1)$ to twistorial oscillators transforming covariantly with respect  to the Lorentz group $SL(2,\mathbf{C})$ with a definite scale dimension one acts with  the intertwining operator \cite{Gunaydin:1998jc,Chiodaroli:2011pp}
\be
T =e^{\frac{\pi}{4}M_{05}}.
\ee
which intertwines between the compact and  the noncompact pictures
\begin{eqnarray}
\mathcal{M}_{a} T & = &  T L_{a} \nonumber\\
\mathcal{N}_{a} T & = &  T R_{a} \nonumber\\
D T    & = &  T E \label{UL}
\end{eqnarray}
where $L_a$ and $R_a$ denote the generators of $SU(2)_L$ and $SU(2)_R$, respectively.
$M_a$ and $N_a$ are the generators  of $SU(2)_\mathcal{M}$ and $SU(2)_\mathcal{N}$  given by the following linear combinations of the Lorentz group generators $M_{\mu\nu}$
\be
\mathcal{M}_a = -\half \lb\half\epsilon_{abc}M_{bc} + i M_{0a}\rb, \qquad \mathcal{N}_a = -\half \lb\half\epsilon_{abc}M_{bc} - i M_{0a}\rb
\ee
They satisfy
\be
\comm{\mathcal{M}_a}{\mathcal{M}_b} = i \epsilon_{abc} \mathcal{M}_c, \qquad \comm{\mathcal{N}_a}{\mathcal{N}_b} = i \epsilon_{abc} \mathcal{N}_c, \qquad \comm{\mathcal{M}_a}{\mathcal{N}_b} = 0
\ee
where $a,b,..=1,2,3$.

The oscillators that transform covariantly under the compact subgroup $SU(2)_L \times SU(2)_R $ get intertwined into the oscillators that transform covariantly under the Lorentz group $SL(2,\mathbb{C})$ .
More specifically the oscillators $a_i (a^i)$ and $b_i (b^i)$ that transform in the $(1/2,0)$ and $(0,1/2) $ representation of $SU(2)_L\times SU(2)_R$  go over to  covariant oscillators transforming as Weyl spinors $(1/2,0)$ and $(0,1/2)$ of the Lorentz group $SL(2,\mathbb{C})$. Denoting the components of the Weyl spinors with undotted ($\a,\b,\ldots =1,2$) and dotted Greek indices ($\ad,\bd,\ldots =1,2 $) one finds :
\bea
\eta^\alpha &= &T a_i T^{-1} = \frac{1}{\sqrt{2}} ( b_i - a^i ) \nn
\lambda_{\alpha} &= &T a^i T^{-1} = \frac{1}{\sqrt{2}} ( b^i + a_i ) \nn
\tilde{\eta}^{\dot{\alpha}}&=& T b_i T^{-1} = \frac{1}{\sqrt{2}} ( a_i - b^i ) \nn
\tilde{\lambda}_{\dot{\alpha}}&=& T b^i T^{-1} = \frac{1}{\sqrt{2}} ( a^i + b_i )
\eea
where $\alpha, \beta, \dot{\alpha}, \dot{\beta},..=1,2$ and the covariant indices on the left hand side match the indices $i,j..$ on the right hand side of the equations above.  They satisfy
\bea
[\eta^\alpha , \lambda_\beta]= \delta_\beta^\alpha \nn
{[} \tilde{\eta}^{\dot{\alpha}} , \tilde{\lambda}_{\dot{\beta}} {]} =\delta^{\dot{\alpha}}_{\dot{\beta}}
\eea
They lead to the standard twistor relations\footnote{ Note the overall minus sign in these expressions compared to \cite{Chiodaroli:2011pp}. This is due to the fact that we are using a mostly positive metric in this paper.}.
\bea
P_{\a\bd}= -( \sigma^\mu P_\mu ) _{\alpha \dot{\beta}} = 2\lambda_\alpha \tilde{\lambda}_{\dot{\beta}} = T a^i b^r T^{-1} \\
K^{\ad \b}= -( \bar{\sigma} ^\mu K_\mu ) ^{\ad \b} = 2\tilde{\eta}^\ad \eta^\b = T a_i b_r T^{-1}
\eea
The dilatation generator in terms of covariant  twistorial oscillators  takes the form:
\be
\Delta = \frac{i}{2} \lb \lambda_\alpha \eta^\alpha +  \tilde{\eta}^{\dot{\alpha}} \tilde{\lambda}_{\dot{\alpha}} \rb
\ee

The Lorentz generators $M_{\mu\nu}$ in a spinorial basis can also be written as bilinears of Lorentz covariant twistorial oscillators:
\be
M_\alpha^{\,\,\,\beta} =  -\frac{i}{2}\lb \sigma^\mu \bar\sigma^\nu \rb_\alpha^{\,\,\,\beta} M_{\mu\nu} =  \lambda_\alpha \eta^\beta - \half \delta_\alpha^{\,\,\,\beta} \lambda_\gamma \eta^\gamma
\ee

\be
\bar{M}_{\,\,\,\dot{\beta}}^{\dot{\alpha}} = -\frac{i}{2} \lb \bar{\sigma}^\mu \sigma^\nu \rb_{\,\,\,\dot{\alpha}}^{\dot{\beta}} M_{\mu\nu} = -\lb \tilde{\eta}^{\dot{\alpha}} \tilde{\lambda}_{\dot{\beta}} - \half \delta_{\,\,\,\dot{\beta}}^{\dot{\alpha}} \tilde{\eta}^{\dot{\gamma}} \tilde{\lambda}_{\dot{\gamma}} \rb
\ee

In this basis the conformal algebra becomes:
\begin{align} \label{Lorentzcomms}
\comm{M_\alpha^{\,\,\,\beta}}{M_\gamma^{\,\,\,\delta}} &=  \delta_\gamma^{\,\,\,\beta} M_\alpha^{\,\,\,\delta} - \delta_\alpha^{\,\,\,\delta} M_\gamma^{\,\,\,\beta},  \qquad
\comm{\bar{M}_{\,\,\,\dot{\beta}}^{\dot{\alpha}}}{\bar{M}_{\,\,\,\dot{\delta}}^{\dot{\gamma}}} =  \delta_{\,\,\,\dot{\beta}}^{\dot{\gamma}} \bar{M}_{\,\,\,\dot{\delta}}^{\dot{\alpha}} -\delta_{\,\,\,\dot{\delta}}^{\dot{\alpha}} \bar{M}_{\,\,\,\dot{\beta}}^{\dot{\gamma}}  \\
\comm{P_{\a \bd}}{M_\g^{\s \d}} &= -\d_\a^\d P_{\g \bd} + \frac{1}{2}\d_\g^\d P_{\a \bd}, \qquad
\comm{P_{\a \bd}}{\bar{M}^\gd_{\s \dd}} = \d_\bd^\gd P_{\a \dd} - \frac{1}{2}\d_\dd^\gd P_{\a \bd} \\
\comm{K^{\ad \b}}{M_\g^{\s \d}} &= \d_\g^\b K^{\ad \d} - \frac{1}{2}\d_\g^\d K^{\ad \b}, \qquad
\comm{K^{\ad \b}}{\bar{M}^\gd_{\s \dd}} = -\d^\ad_\dd K^{\gd \b} + \frac{1}{2}\d_\dd^\gd K^{\ad \b} \\
\comm{\Delta}{K^{\ad \b}} &= -iK^{\ad \b}, \quad \comm{\Delta}{M_\a^{\s \b}}=\comm{\Delta}{\bar{M}^\ad_{\s \bd}}=0, \quad \comm{\Delta}{P_{\a \bd}} = iP_{\a \bd} \\
\comm{P_{\a\bd}}{K^{\gd\d}} & = 4 \lb \d_\a^\d \bar{M}^\gd_{\s \bd} - \d_\bd^\gd M_\a^{\s \d} + i \Delta \rb
\end{align}

The linear Casimir operator  $\mathcal{Z}= N_a - N_b= a^ia_i -b^r b_r$ when expressed in terms of $SL(2,\mathbb{C})$ covariant oscillators becomes
\be
\mathcal{Z}= N_a - N_b = \tilde{\lambda}_{\dot{\alpha}} \tilde{\eta}^{\dot{\alpha}} - \lambda_\alpha \eta^\alpha
\ee
which shows that $\frac{1}{2} \mathcal{Z}$ is the helicity operator.

Denoting the lowest energy irreps in the compact basis as
 $|\Omega (j_L,j_R,E )\rangle$  one can show  that the coherent states of the form
\be
e^{-i x^{\mu}P_{\mu}}T|\Omega (j_L,j_R,E )\rangle \equiv |\Phi^{\ell}_{j_\mathcal{M},j_\mathcal{N}}(x_\mu) \rangle
\ee
 transform exactly like the states created by the action of  conformal fields $\Phi^{\ell}_{j_\mathcal{M},j_\mathcal{N}}(x_\mu) $ acting on the vacuum vector $| 0\rangle $
 \[ \Phi^{\ell}_{j_\mathcal{M},j_\mathcal{N}}(x^{\mu}) | 0 \rangle  \cong |\Phi^{\ell}_{j_\mathcal{M},j_\mathcal{N}}(x_\mu) \rangle\]
with exact numerical coincidence of the compact
and the covariant labels $(j_{L},j_{R},E)$ and $(j_{\mathcal{M}},j_{\mathcal{N}},-l)$, respectively, where $l$ is the scale dimension \cite{Gunaydin:1998jc}.
The doubletons  correspond to massless conformal fields transforming in the
$(j_L,j_R)$ representation of the Lorentz group $SL(2,\mathbb{C})$  whose conformal (scaling
dimension) is $\ell=-E$ where $E$ is the eigenvalue of the $U(1)$ generator
which is the conformal Hamiltonian (or $AdS_5$ energy) \cite{Gunaydin:1984fk,Gunaydin:1998jc,Gunaydin:1998sw}.

\subsection{Quasiconformal approach to the minimal unitary representation of $SO(4,2)$ and its deformations \label{qcminrep}}
\label{sect4dqc}
The construction of the minimal unitary representation (minrep) of the $4d$ conformal group $SO(4,2)$ by quantization of its quasiconformal realization and its  deformations were given in  \cite{Fernando:2009fq}, which we shall reformulate in this section in terms of what we call deformed twistorial oscillators which transform nonlinearly under the Lorentz group.

The group $SO(4,2)$ can be realized as a quasiconformal group that leaves invariant light-like separations with respect to a quartic distance function in five dimensions. The quantization of this geometric action leads to a nonlinear realization of the generators of $SO(4,2)$ in terms of a singlet coordinate  $x$, its conjugate momentum $p$ and two ordinary bosonic oscillators  $d,d^\dagger$ and $g,g^\dagger$ satisfying \cite{Fernando:2009fq}:
\be
\comm{x}{p} =i, \qquad\qquad \comm{d}{d^\dagger} =1, \qquad\qquad \comm{g}{g^\dagger} =1
\ee
The nonlinearities can be absorbed into certain  ``singular" oscillators which are functions of the coordinate $x$, momentum $p$  and the oscillators $d,g,d^\dagger,g^\dagger$:
\begin{equation}
A_{\mathcal{L}} = a - \frac{\mathcal{L}}{\sqrt{2} \, x}
\qquad \qquad \qquad
A_{\mathcal{L}}^\dag = a^\dag - \frac{\mathcal{L}}{\sqrt{2} \, x}
\end{equation}
where
\bea
a \eq \frac{1}{\sqrt{2}} ( x + i p) \nn
a^\dagger \eq \frac{1}{\sqrt{2}} ( x - i p) \nn
\mathcal{L} \eq N_d - N_g - \frac{1}{2}= d^\dagger d - g^\dagger g - \frac{1}{2}
\eea
They satisfy the following commutation relations:
\begin{equation}
\begin{split}
\comm{A_{\mathcal{G}}}{A_{\mathcal{K}}}
 &= - \frac{\left( \mathcal{G} - \mathcal{K} \right)}{2 \, x^2} \\
\comm{A_{\mathcal{G}}^\dag}{A_{\mathcal{K}}^\dag}
 &= + \frac{\left( \mathcal{G} - \mathcal{K} \right)}{2 \, x^2} \\
\comm{A_{\mathcal{G}}}{A_{\mathcal{K}}^\dag}
 &= 1 + \frac{\left( \mathcal{G} + \mathcal{K} \right)}{2 \, x^2}
\end{split}
\end{equation}
assuming that $[\mathcal{G}, \mathcal{K}]=0$.

The realization of the minrep of $SO(4,2)$ obtained by the quasiconformal approach  is  nonlinear and ``interacting" in the sense that they involve operators that are cubic or quartic in terms of the oscillators in contrast to the covariant twistorial oscillator realization, reviewed in section \ref{4d-doubleton-section} \cite{Gunaydin:1984fk}, which involves only bilinears.
The algebra $\mathfrak{so}(4,2)$ can  be given a 3-graded decomposition with respect to the conformal Hamiltonian, which is  referred to as the compact 3-grading and the generators in this basis are reproduced in Appendix \ref{app-c3grading} following  \cite{Fernando:2009fq}.

The Lie algebra of $SO(4,2)$  has also  a noncompact (conformal) three graded decomposition  determined by the dilatation generator $\Delta$ as well
\bea
\mathfrak{so}(4,2) \eq \mathfrak{N}^- \oplus \mathfrak{N}^0  \oplus \mathfrak{N}^+ \\
\eq K_\mu \oplus (M_{\mu\nu} \oplus \Delta) \oplus P_\mu
\eea
One can  write the generators of quantized  quasiconformal action  of $SO(4,2)$ as bilinears  of  deformed twistorial oscillators $Z^\alpha, \widetilde{Z}^{\dot{\alpha}}, Y_\alpha, \widetilde{Y}_{\dot{\alpha}}$ ($\alpha, \dot{\alpha}=1,2$) which are defined as:

\begin{align}
Z_1 &= \frac{A_\ml}{\sqrt2} - i \,g^\dag, \qquad  Y^1 = -\frac{A_\ml^\dag}{\sqrt2} + i \,g \\
\widetilde{Z}_{\dot{1}} &= \frac{A^\dag_\ml}{\sqrt2} + i\, g, \qquad \widetilde{Y}^{\dot{1}} = \frac{A_\ml}{\sqrt2} + i\, g^\dag \\
Z_2 &= -\frac{A^\dag_{-\ml}}{\sqrt2} - i \,d, \qquad Y^2 = -\frac{A_{-\ml}}{\sqrt2} - i \,d^\dag \\
\widetilde{Z}_{\dot{2}} &= - \frac{A_{-\ml}}{\sqrt2} + i\, d^\dag, \qquad \widetilde{Y}^{\dot{2}} = \frac{A^\dag_{-\ml}}{\sqrt2} - i\, d
\end{align}
Using $\lb \sigma^\mu \rb_{\alpha \dot{\alpha}} = (\mathbbm{1}_2,\vec{\sigma})$ and $\lb \bar\sigma^\mu \rb^{\dot{\alpha}\alpha} =  (-\mathbbm{1}_2,\vec{\sigma})$ one finds that  the generators of translations and special conformal transformations can be written  as follows\footnote{Note that in our conventions $P^0$ is positive definite.}:
\begin{align}
P_{\alpha \dot{\beta}} &= \lb \sigma^\mu P_\mu \rb_{\alpha\dot{\beta}} = -Z_\alpha \widetilde{Z}_{\dot{\beta}} \\
K^{\dot{\alpha} \beta} &= \lb \bar{\sigma}^\mu K_\mu \rb^{\dot{\alpha}\beta} = -\widetilde{Y}^{\dot{\alpha}} Y^{\beta}
\end{align}
We see that the operators  $Z$ and $Y$ in the quasiconformal realization play similar roles as covariant  twistorial oscillators $\lambda$ and $\eta$ in  the doubleton realization. However, they transform nonlinearly under the Lorentz group and  their commutations relations  are given in Appendix \ref{app-yztwist}

The dilatation generator in terms of deformed twistorial oscillators takes the form:
\be
\Delta = \frac{i}{4} \lb Z_\alpha Y^\alpha +  \widetilde{Y}^{\dot{\alpha}} \widetilde{Z}_{\dot{\alpha}} \rb
\ee
The Lorentz group generators $M_{\mu\nu}$ in a spinorial basis can also be written as bilinears of these deformed  twistorial oscillators:
\be
M_\alpha^{\,\,\,\beta} =  -\frac{i}{2}\lb \sigma^\mu \bar\sigma^\nu \rb_\alpha^{\,\,\,\beta} M_{\mu\nu} = \half \lb Z_\alpha Y^\beta - \half \delta_\alpha^{\,\,\,\beta} Z_\gamma Y^\gamma \rb
\ee

\be
\bar{M}_{\,\,\,\dot{\beta}}^{\dot{\alpha}} = -\frac{i}{2} \lb \bar{\sigma}^\mu \sigma^\nu \rb_{\,\,\,\dot{\alpha}}^{\dot{\beta}} M_{\mu\nu} = -\half \lb \widetilde{Y}^{\dot{\alpha}} \widetilde{Z}_{\dot{\beta}} - \half \delta_{\,\,\,\dot{\beta}}^{\dot{\alpha}} \widetilde{Y}^{\dot{\gamma}} \widetilde{Z}_{\dot{\gamma}} \rb
\ee
We should stress the important point that even though the deformed twistorial oscillators transform nonlinearly under the Lorentz group, their bilinears $P_{\alpha \dot{\beta}}, K^{\dot{\alpha} \beta},M_\alpha^{\,\,\,\beta} $ and $\bar{M}_{\,\,\,\dot{\beta}}^{\dot{\alpha}}$ transform covariantly and satisfy the commutation relations given in  equations \ref{Lorentzcomms}.
\subsection{Deformations of the minimal unitary representation of $SU(2,2)$}
\label{sect4dzeta}

As was shown in \cite{Fernando:2009fq}, the minimal unitary representation of $SU(2,2)$  that corresponds to a conformal scalar field admits a one-parameter, $\zeta$,   family of deformations that correspond to massless conformal fields of helicity $\frac{\zeta}{2}$ in four dimensions, which can be continuous. For non-integer values of the deformation parameter $\zeta$ they correspond, in general, to unitary representations of an infinite covering of the conformal group.


The generators of the deformed minrep take the same form as given in section \ref{sect4dqc} with the simple replacement of  the singular oscillators $A_{\mathcal{L}}$ and $A_{\mathcal{L}}^\dagger$ by  ``deformed" singular oscillators:
\begin{equation}
A_{\mathcal{L}_\zeta} = a - \frac{\mathcal{L}_\zeta}{\sqrt{2} \, x}
\qquad \qquad \qquad
A_{\mathcal{L}_\zeta}^\dag = a^\dag - \frac{\mathcal{L}_\zeta}{\sqrt{2} \, x}
\end{equation}
where
\be
\ml_\zeta = \ml + \zeta = N_d - N_g + \zeta - \half
\ee
Since $\zeta/2$ labels the helicity we define ``helicity deformed twistors"  as  follows:
\begin{align} \label{deformedtwistors}
Z_1(\zeta) &= \frac{A_{\mathcal{L}_\zeta}}{\sqrt2} - i \,g^\dag, \qquad  Y^1(\zeta) = -\frac{A_{\mathcal{L}_\zeta}^\dag}{\sqrt2} + i \,g \\
\widetilde{Z}_{\dot{1}}(\zeta) &= \frac{A^\dag_{\mathcal{L}_\zeta}}{\sqrt2} + i\, g, \qquad \widetilde{Y}^{\dot{1}}(\zeta) = \frac{A_{\mathcal{L}_\zeta}}{\sqrt2} + i\, g^\dag \\
Z_2(\zeta) &= -\frac{A^\dag_{-{\mathcal{L}_\zeta}}}{\sqrt2} - i \,d, \qquad Y^2(\zeta) = -\frac{A_{-{\mathcal{L}_\zeta}}}{\sqrt2} - i \,d^\dag \\
\widetilde{Z}_{\dot{2}}(\zeta) &= - \frac{A_{-{\mathcal{L}_\zeta}}}{\sqrt2} + i\, d^\dag, \qquad \widetilde{Y}^{\dot{2}}(\zeta) = \frac{A^\dag_{-{\mathcal{L}_\zeta}}}{\sqrt2} - i\, d
\end{align}
The realization of the minimal unitary representation in terms of deformed  twistors carry over to realization in terms of helicity deformed twistors:
\begin{align}
P_{\alpha \dot{\beta}} &= \lb \sigma^\mu P_\mu \rb_{\alpha\dot{\beta}} = -  Z_\alpha(\zeta)  \widetilde{Z}_{\dot{\beta}}(\zeta) \\
K^{\dot{\alpha} \beta} &= \lb \bar{\sigma}^\mu K_\mu \rb^{\dot{\alpha}\beta} = - \widetilde{Y}^{\dot{\alpha}}(\zeta)  Y^\beta (\zeta)
\end{align}
The dilatation generator then  takes the form:
\be
\Delta = \frac{i}{4} \lb Z_\alpha (\zeta) Y^\alpha (\zeta) +  \widetilde{Y}^{\dot{\alpha}}(\zeta) \widetilde{Z}_{\dot{\alpha}}(\zeta) \rb
\ee
and the Lorentz generators $M_{\mu\nu}$ also take the same form in terms of helicity deformed twistors:
\be
M_\alpha^{\,\,\,\beta} =  -\frac{i}{2} \lb \sigma^\mu \bar\sigma^\nu \rb_\alpha^{\,\,\,\beta} M_{\mu\nu} = \half \lb Z_\alpha(\zeta) Y^\beta(\zeta) - \half \delta_\alpha^{\,\,\,\beta} Z_\gamma (\zeta) Y^\gamma (\zeta) \rb
\ee

\be
\bar{M}_{\,\,\,\dot{\beta}}^{\dot{\alpha}} = -\frac{i}{2} \lb \bar{\sigma}^\mu \sigma^\nu \rb_{\,\,\,\dot{\alpha}}^{\dot{\beta}} M_{\mu\nu} = -\half \lb \widetilde{Y}^{\dot{\alpha}}(\zeta) \widetilde{Z}_{\dot{\beta}}(\zeta) - \half \delta_{\,\,\,\dot{\beta}}^{\dot{\alpha}} \widetilde{Y}^{\dot{\gamma}}(\zeta) \widetilde{Z}_{\dot{\gamma}}(\zeta) \rb
\ee
The realization of the generators of $SU(2,2)$ in terms of helicity deformed twistors describe positive energy unitary irreducible representations which can best be seen by going over to the compact three-grading reviewed in Appendix  \ref{app-c3grading}.

Since the quasiconformal realization of the minrep and its deformations are nonlinear the tensoring procedure in quasiconformal framework is a non-trivial and open problem. However since the representations with integer values of $\zeta$ are isomorphic to the doubleton representations, the result of tensoring is already known and was discussed in section \ref{4d-doubleton-section}.

\subsection{Minimal unitary supermultiplet of $SU(2,2|4)$, its deformations and  deformed  twistors }
\label{section-superconformal}
The construction of the minimal unitary representations of noncompact Lie algebras  by quantization of their quasiconformal realizations extends to noncompact Lie superalgebras \cite{Gunaydin:2006vz,Fernando:2009fq,Fernando:2010dp,Fernando:2010ia}. In particular, the minimal unitary supermultiplets of $SU(2,2|N)$ and their deformations were studied  in \cite{Fernando:2009fq} using quasiconformal methods.
In this section we shall reformulate the minimal unitary realization of  $4d$ superconformal algebra  $SU(2,2|4)$  and its deformations in terms of deformed twistorial oscillators \footnote{ $PSU(2,2|4)$ is the symmetry superalgebra of $IIB$ supergravity compactified over $AdS_5\times S^5$ ymmetry \cite{Gunaydin:1984fk}.}.

The superconformal algebra $\mathfrak{su}(2,2|4)$  can be given a (noncompact) 5-graded decomposition  with respect to the dilatation generator $\Delta$:
\bea
\mathfrak{su}(2,2|4) \eq \mathfrak{N}^{-1} \oplus \mathfrak{N}^{-1/2} \oplus \mathfrak{N}^0  \oplus \mathfrak{N}^{+1/2} \oplus \mathfrak{N}^{+1} \\
\eq K^{\ad\b} \oplus S_{I}^{\,\,\,\alpha}, \bar{S}^{I\dot{\alpha}} \oplus (M_\a^{\s\b} \oplus \bar{M}^\ad_{\s\bd} \oplus \Delta \oplus R^I_{\,\,\, J}) \oplus Q^I_{\,\,\,\alpha}, \bar{Q}_{I\dot{\alpha}} \oplus P_{\a\bd}, \nn
&& \qquad \qquad\qquad\qquad\qquad\qquad\qquad\qquad\qquad \qquad (I,J=1,2,3,4) \nonumber
\eea
where the grade zero subspace $ \mathfrak{N}^0 $ consists of the Lorentz algebra $\mathfrak{so}(3,1)$ ($M_\a^{\s\b}, \bar{M}^\ad_{\s\bd}$), the dilatations ($\Delta$) and R-symmetry $\mathfrak{su}(4)$ ($R^I_{\,\,\, J}$) generators, grade $+1$ and $-1$ subspaces consist of translation ($P_{\a\bd}$) and special conformal generators ($K^{\ad\b}$) and the grade +1/2 and -1/2 subspaces consist of Poincar\'e supersymmetry ($ Q^I_{\,\,\,\alpha}, \bar{Q}_{I\dot{\alpha}}$) and special conformal  supersymmetry generators ($S_{I}^{\,\,\,\alpha}, \bar{S}^{I\dot{\alpha}}$) respectively.

The helicity deformed twistors for the superalgebra $SU(2,2|4)$ are obtained from the deformed twistors of $SU(2,2)$  by replacing $\mathcal{L}_\zeta$  in the corresponding  deformed singular oscillators $A_{\ml_{\zeta}} $ with $\mathcal{L}^s_\zeta$  :
\be
\mathcal{L}_{\zeta} \longrightarrow \ml_{\zeta}^s = N_d - N_g + N_\xi + \zeta - \frac{5}{2}
\ee
where  $N_\xi = \xi^J \xi_J$ is the number operator of four fermionic oscillators
$\xi_I \, ( \xi^J) $ ($I,J = 1,2,3,4$) that satisfy
\be
\acomm{\xi_I}{\xi^J} = \delta_I^J
\ee

The expressions for the generators of $SU(2,2)$ given in \ref{sect4dqc} get modified as follows in going over to $SU(2,2|4)$:
\begin{align}
P_{\alpha \dot{\beta}} &=  -Z_\alpha^s (\zeta) \widetilde{Z}_{\dot{\beta}}^s (\zeta) \\
K^{\dot{\alpha} \beta} &= -\widetilde{Y}^{s\dot{\alpha}} (\zeta) Y^{s\beta} (\zeta)
\end{align}
\be
\Delta = \frac{i}{4} \lb Z_\alpha^s (\zeta) Y^{s\alpha} (\zeta) +  \widetilde{Y}^{s\dot{\alpha}} (\zeta) \widetilde{Z}_{\dot{\alpha}}^s (\zeta) \rb
\ee
\be
M_\alpha^{\,\,\,\beta} =  \half \lb Z_\alpha^s (\zeta) Y^{s \beta}(\zeta) - \half \delta_\alpha^{\,\,\,\beta} Z_\gamma^s (\zeta) Y^{s \gamma} (\zeta) \rb
\ee
\be
\bar{M}_{\,\,\,\dot{\beta}}^{\dot{\alpha}}  = -\half \lb \widetilde{Y}^{s\dot{\alpha}}(\zeta) \widetilde{Z}_{\dot{\beta}}^s (\zeta) - \half \delta_{\,\,\,\dot{\beta}}^{\dot{\alpha}} \widetilde{Y}^{s\dot{\gamma}}(\zeta) \widetilde{Z}_{\dot{\gamma}}^s (\zeta) \rb
\ee
where the "supersymmetric" helicity deformed twistors are defined as:
\begin{align} \label{deformedtwistors}
Z^s_1(\zeta) &= \frac{A_{\mathcal{L}^s_\zeta}}{\sqrt2} - i \,g^\dag, \qquad  Y^{s1}(\zeta) = -\frac{A_{\mathcal{L}^s_\zeta}^\dag}{\sqrt2} + i \,g  \nn
\widetilde{Z}^s_{\dot{1}}(\zeta) &= \frac{A^\dag_{\mathcal{L}^s_\zeta}}{\sqrt2} + i\, g, \qquad \widetilde{Y}^{s\dot{1}}(\zeta) = \frac{A_{\mathcal{L}^s_\zeta}}{\sqrt2} + i\, g^\dag \nn
Z^s_2(\zeta) &= -\frac{A^\dag_{-{\mathcal{L}^s_\zeta}}}{\sqrt2} - i \,d, \qquad Y^{s2}(\zeta) = -\frac{A_{-{\mathcal{L}^s_\zeta}}}{\sqrt2} - i \,d^\dag \nn
\widetilde{Z}^s_{\dot{2}}(\zeta) &= - \frac{A_{-{\mathcal{L}^s_\zeta}}}{\sqrt2} + i\, d^\dag, \qquad \widetilde{Y}^{s\dot{2}}(\zeta) = \frac{A^\dag_{-{\mathcal{L}^s_\zeta}}}{\sqrt2} - i\, d \nonumber
\end{align}

The supersymmetry  generators of $SU(2,2|4)$ are given by the bilinears of   deformed twistorial oscillators  and fermionic oscillators:
\begin{align}
Q^I_{\,\,\,\alpha} &= Z_\alpha^s (\zeta) \xi^I, \qquad \bar{Q}_{I\dot{\alpha}} = -\xi_I \widetilde{Z}_{\dot{\alpha}}^s(\zeta) \\
S_I^{\,\,\,\alpha} &= -\xi_I Y^{s \alpha }(\zeta), \qquad \bar{S}^{I\dot{\alpha}} = \widetilde{Y}^{s\dot{ \alpha} }(\zeta) \xi^I \label{QSsusy}
\end{align}
The  generators $R^I_{\,\,\, J}$ of R-symmetry group $SU(4)$  are given by:
\be
R^I_{\,\,\, J} = \xi^I \xi_J - \frac{1}{4}\delta^I_{\,\,\, J} \xi^K \xi_K
\ee
They satisfy  the following anti-commutation relations:
\bea
\acomm{Q^I_{\,\,\,\alpha}}{\bar{Q}_{J\dot{\beta}}} \eq \delta^I_J P_{\alpha \dot{\beta}} \\
\acomm{\bar S^{I\dot{\alpha}}}{S_J^{\,\,\,\beta}} \eq \delta^I_J K^{\dot{\alpha} \beta}
\eea
\bea
\acomm{Q^I_{\,\,\,\alpha}}{S_J^{\,\,\,\beta}} \eq -2 \delta^I_J M_\a^{\s \b} + 2\delta^\beta_\alpha R^I_{\,\,\,J} + \delta^I_J \delta^\beta_\alpha \lb i\Delta + C \rb \\
\acomm{\bar S^{I\dot{\alpha}}}{\bar{Q}_{J\dot{\beta}}} \eq 2 \delta^I_J \bar{M}^{\dot{\alpha}}_{\,\,\,\dot{\beta}} - 2\delta^{\dot{\beta}}_{\dot{\alpha}} R^I_{\,\,\,J} + \delta^I_J \delta^{\dot{\beta}}_{\dot{\alpha}} \lb i\Delta - C \rb
\eea
where $C=\frac{\zeta}{2}$ is the central charge.

The commutators  of conformal group generators with  supersymmetry generators are as follows:
\begin{align}
\comm{P_{\a \bd}}{S_I^{\s \g}} &= 2 \d_\a^\g \bar{Q}_{I\bd}, \qquad \comm{K^{\ad \b}}{Q_\g^{\s I}} = -2 \d_\g^\b \bar{S}^{I\ad} \\
\comm{P_{\a \bd}}{\bar{S}^{I^ \gd}} &= 2 \d_\bd^\gd Q_\a^{\s I}, \qquad \comm{K^{\ad \b}}{\bar{Q}_{I\gd}} = -2 \d_\gd^\ad S_I^{\s \b}
\end{align}
\begin{align}
\comm{M_\alpha^{\,\,\,\beta}}{Q^I_{\,\,\,\gamma}} & = \delta^\beta_\gamma Q^I_{\,\,\,\alpha} - \frac{1}{2} \delta_\alpha^\beta Q^I_{\,\,\,\gamma}, \qquad \comm{M_\alpha^{\,\,\,\beta}}{S_I^{\,\,\,\gamma}}  =  -\delta^\alpha_\gamma S_I^{\,\,\,\beta} + \frac{1}{2} \delta_\alpha^\beta S_I^{\,\,\,\gamma}  \\
\comm{\bar{M}^{\dot{\alpha}}_{\,\,\,\dot{\beta}}}{\bar{Q}_{I\dot{\gamma}}} & = -\delta^{\dot{\alpha}}_{\dot{\gamma}} \bar{Q}_{I\dot{\beta}} + \frac{1}{2} \delta^{\dot{\alpha}}_{\dot{\beta}} \bar{Q}_{I\dot{\gamma}}, \qquad \comm{\bar{M}^{\dot{\alpha}}_{\,\,\,\dot{\beta}}}{\bar{S}^{I\dot{\gamma}}}  = \delta^{\dot{\gamma}}_{\dot{\beta}} \bar{S}^{I\dot{\alpha}} - \frac{1}{2} \delta^{\dot{\alpha}}_{\dot{\beta}} \bar{Q}^{I\dot{\gamma}}
\end{align}
\begin{align}
\comm{\Delta}{Q^I_{\,\,\,\alpha}} &= \frac{i}{2}Q^I_{\,\,\,\alpha}, \qquad \comm{\Delta}{ \bar{Q}_{I\dot{\alpha}}} =  \frac{i}{2}\bar{Q}_{I\dot{\alpha}} \\
\comm{\Delta}{S_I^{\,\,\,\alpha}} &= -\frac{i}{2}S_I^{\,\,\,\alpha}, \qquad \comm{\Delta}{ \bar{S}^{I\dot{\alpha}}} = - \frac{i}{2}\bar{S}^{I\dot{\alpha}}
\end{align}

The $\mathfrak{su}(4)_R$ generators satisfy the following commutation relations:
\be
\comm{R^I_{\,\,\, J}}{R^K_{\,\,\, L}} = \delta^K_J R^I_{\,\,\, L} - \delta^I_L R^K_{\,\,\, J}
\ee
They act on the R-symmetry indices $I,J$ of  the supersymmetry generators as follows:
\begin{align}
\comm{R^I_{\,\,\, J}}{Q^K_{\,\,\,\alpha}} &= \delta^K_J Q^I_{\,\,\,\alpha} - \frac{1}{4}\delta^I_J Q^K_{\,\,\,\alpha}, \qquad \comm{R^I_{\,\,\, J}}{\bar{Q}_{K\dot{\alpha}}} = -\delta^I_K \bar{Q}_{J\dot{\alpha}} + \frac{1}{4}\delta^I_J \bar{Q}_{K\dot{\alpha}} \\
\comm{R^I_{\,\,\, J}}{S_K^{\,\,\,\alpha}} &= -\delta^I_K S_J^{\,\,\,\alpha} + \frac{1}{4}\delta^I_J S_K^{\,\,\,\alpha}, \qquad \comm{R^I_{\,\,\, J}}{\bar{S}^{K\dot{\alpha}}} = \delta^K_J \bar{S}^{I\dot{\alpha}} - \frac{1}{4}\delta^I_J \bar{S}^{K\dot{\alpha}}
\end{align}
The minimal unitary representation of $PSU(2,2|4)$ is obtained when the deformation parameter $\zeta$, which is also the central charge, vanishes. The resulting minimal unitary supermultiplet of massless conformal fields in $d=4$ is simply the $N=4$ Yang-Mills supermultiplet \cite{Fernando:2009fq}.  For each value of the deformation parameter $\zeta$ one obtains an irreducible unitary representation of $SU(2,2|4)$. For integer values of the deformation parameter these unitary representations are isomorphic to doubleton supermultiplets studied in \cite{Gunaydin:1984fk,Gunaydin:1998jc,Gunaydin:1998sw}.

The unitarity of the representations of $SU(2,2|N)$ may not be manifest in the Lorentz covariant noncompact five grading. It is however manifestly unitary in compact three grading with respect to the subsupergroup $SU(2|N-M)\times SU(2|M) \times U(1)$ as was shown for the the doubletons in \cite{Gunaydin:1984fk,Gunaydin:1998sw} and for the quasiconformal construction in \cite{Fernando:2009fq}.   The Lie algebra of $SU(4)$ can be given a 3-graded structure with respect to the Lie algebra of its subgroup $SU(2)\times SU(2)\times U(1)$. Similarly the Lie superalgebra $SU(2,2|4)$ can be given a 3-graded decomposition with respect to its subalgebra $SU(2|2)\times SU(2|2)\times U(1)$. This is the basis that was originally used by Gunaydin and Marcus \cite{Gunaydin:1984fk}  in constructing the spectrum of $IIB$ supergravity over $AdS_5\times S^5$ using twistorial oscillators. In this basis, choosing the Fock vacuum as the lowest  weight vector leads to CPT-self-conjugate supermultiplets and it is also the preferred basis in applications to  integrable spin chains. The corresponding compact 3-grading of
the quasiconformal realization of $SU(2,2|4)$ was given in \cite{Fernando:2009fq}, to which we refer for details.


\section{Higher spin (super-)algebras, Joseph ideals and their deformations}
\label{sect-joseph}
In this section we start by reviewing Eastwood's results \cite{Eastwood:2002su,eastwood2005uniqueness} on defining $HS(\mfg)$ algebras as the quotient of universal enveloping algebra $\mathscr{U}(\mfg)$ by its Joseph ideal $\mathscr{J}(\mfg)$. We will then explicitly compute the Joseph ideal for $SO(3,2)$, $SO(4,2)$ and its deformations, using the Eastwood formula \cite{eastwood2005uniqueness} and recast it in a Lorentz covariant form.

The universal enveloping algebra $\mathscr{U}(\mfg)$, $\mfg = \mathfrak{so}(d-1,2)$ is defined as follows:
\be
\mathscr{U}(\mfg) = \mathscr{G}/\mathscr{I}
\ee
where $\mathscr{G}$ is the associative algebra freely generated by elements of $\mfg$, and $\mathscr{I}$ is the ideal of $\mathscr{G}$ generated by elements of form $g h-h g - \comm{g}{h} (g,h \in \mfg)$.

The enveloping algebra $\mathscr{U}(\mathfrak{g})$ can be decomposed into standard adjoint action of $\mathfrak{g}$ which by Poincare-Birkhoff-Witt theorem is equivalent to computing symmetric products $M_{AB}\sim\parbox{10pt}{\YoungAA}\,$. In particular, $\bigotimes^2 \mathfrak{so}(d-1,2)$ decomposes as:
\be
\parbox{10pt}{\YoungAA} \otimes \parbox{10pt}{\YoungAA} = \parbox{20pt}{\YoungBB}\oplus\parbox{10pt}
{\YoungAAAA}\oplus\parbox{20pt}{\YoungB}\oplus\bullet
\label{ytdecomp}
\ee
where $\bullet$ is the quadratic Casimir $C_2 \sim M^{~A}_B M^{~B}_A$. It was already noted in \cite{Vasiliev:1999ba} that the higher spin algebra $HS(\mfg)$ must be a quotient of $\mathscr{U}(\mfg)$ because the higher spin fields in $AdS_d$ are described by traceless two row Young tableaux. Thus the relevant ideal should quotient out  all the diagrams except the first one in the above decomposition. This ideal was identified in \cite{Eastwood:2002su} to be the Joseph ideal or the annihilator of the minimal unitary representation (scalar doubleton). The uniqueness of this quadratic ideal in $\mathscr{U}(\mfg)$ was proved in \cite{eastwood2005uniqueness} and an explicit formula for the generator $J_{ABCD}$  of the ideal was given as :
\bea
J_{ABCD} &=& M_{AB}M_{CD} - M_{AB}\circledcirc M_{CD} - \frac{1}{2}\comm{M_{AB}}{M_{CD}} + \frac{n-4}{4(n-1)(n-2)} \langle M_{AB},M_{CD}  \rangle \, \mathbf{1} \nonumber  \\
&=& \frac{1}{2}  M_{AB} \cdot M_{CD} - M_{AB}\circledcirc M_{CD} + \frac{n-4}{4(n-1)(n-2)} \langle M_{AB},M_{CD} \rangle \,  \mathbf{1} \label{Joseph}
\eea
where the dot  $
\cdot$ denotes the symmetric product  \be M_{AB}\cdot M_{CD} \equiv M_{AB} M_{CD}+M_{CD}M_{AB} \ee
 of the generators and  $\langle M_{AB},M_{CD} \rangle$ is the Killing form of $SO(n-2,2)$.
 $\eta_{AB}$ is the $SO(n-2,2)$ invariant metric and the symbol $\circledcirc$ denotes the Cartan product of two generators, which  for $SO(n-2,2)$, can be written in the form \cite{eastwood2005cartan}:
\bea
M_{AB}\circledcirc M_{CD} \eq \third M_{AB}M_{CD} + \third M_{DC}M_{BA} + \sixth M_{AC}M_{BD} \nn
&& -\sixth M_{AD}M_{BC} + \sixth M_{DB}M_{CA} - \sixth M_{CB}M_{DA} \nn
&& -\frac{1}{2(n-2)}\left(M_{AE}M_C^E\eta_{BD}-M_{BE}M_C^E\eta_{AD}+M_{BE}M_D^E\eta_{AC}-M_{AE}M_D^E\eta_{BC}\right) \nn
&& -\frac{1}{2(n-2)}\left(M_{CE}M_A^E\eta_{BD}-M_{CE}M_B^E\eta_{AD}+M_{DE}M_B^E\eta_{AC}-M_{DE}M_A^E\delta_{BC}\right) \nn
&& +\frac{1}{(n-1)(n-2)} M_{EF}M^{EF}\left(\eta_{AC}\eta_{BD}-\eta_{BC}\eta_{AD}\right)
\eea

The Killing term is given by
\be
\langle M_{AB},M_{CD} \rangle = h\, M_{EF}M_{GH} (\eta^{EG}\eta^{FH}-\eta^{EH}\eta^{FG}) (\eta_{AC}\eta_{BD}-\eta_{AD}\eta_{BC})
\ee
where $h = \frac{2(n-2)}{n(4-n)}$ is a c-number fixed by requiring that all possible contractions of $J_{ABCD}$ with the  metric vanish.  We shall refer to the operator $J_{ABCD}$ as the generator of the Joseph ideal.

The generator of the Joseph ideal defined in equation \ref{Joseph} contains exactly the operators that correspond to the  ``unwanted" diagrams described in equation \ref{ytdecomp} and quotienting the enveloping algebra by this ideal guarantees that the resulting algebra will contain only two row traceless diagrams and thus correctly describe massless higher spin fields.

In the following sections we will compute the generator $J_{ABCD}$ for $d=3$ and 4 conformal algebras $SO(3,2)$ and $SO(4,2)$ in various realizations discussed in previous sections. We shall also decompose the  generators of the Joseph ideal of $SO(3,2)$ and $SO(4,2)$ with respect to the corresponding Lorentz groups $SO(2,1)$ and $SO(3,1)$, respectvely. The Lorentz  covariant decomposition makes the massless nature of the minimal unitary  representations explicit along with certain other identities that must be satisfied within the representation in order for it to be annihilated by the Joseph ideal. This will also allow us to define the annihilators of the deformations of the minrep of $SO(4,2)$ and the corresponding deformations of the Joseph ideal. These deformations define a one parameter family of $AdS_5/CFT_4$ $HS$ algebras.
\subsection{Joseph ideal for $SO(3,2)$ singletons}
\label{singletons}
We will now use the twistorial oscillator realization for $SO(3,2)$ described in section \ref{so32twist}. For $Sp(4,\mathbb{R})= SO(3,2)$, the generator $J_{ABCD}$ of  the Joseph ideal is
\bea
J_{ABCD} &=& M_{AB}M_{CD} - M_{AB}\circledcirc M_{CD} - \frac{1}{2}\comm{M_{AB}}{M_{CD}} - \frac{1}{40} \langle M_{AB},M_{CD} \rangle \nn
&=& \frac{1}{2} M_{AB} \cdot M_{CD} - M_{AB}\circledcirc M_{CD}  - \frac{1}{40} \langle M_{AB},M_{CD} \rangle
\label{jos3d}
\eea
Substituting the realization of $Sp(4,\mathbb{R})= SO(3,2)$ in terms of a twistorial Majorana spinor $\Psi$ one finds that the operator $J_{ABCD}$ vanishes identically.

Considered as the the three dimensional conformal group the minreps of $Sp(4,\mathbb{R})$ ($Di$  and $Rac$) correspond to   massless scalar and spinor fields   which are known to be the only massless representations of the Poincar\'e group in three dimensions \cite{binegar1991unitarization}.

If instead of a twistorial Majorana spinor one considers a twistorial Dirac spinor corresponding to taking two copies (colors) of the Majorana spinor one finds that the generators $J_{ABCD}$ of the Joseph ideal do not vanish identically and hence they do not correspond to minimal unitary representations. The corresponding Fock space decomposes into an infinite set of irreducible unitary representations of    $ Sp(4,\mathbb{R})$, which correspond to  the massless fields in $AdS_4$ \cite{Angelopoulos:1980wg}. Taking more than two colors in the realization of the Lie algebra of $Sp(4,\mathbb{R})$ as bilinears of oscillators leads to representations corresponding to massive fields in $AdS_4$  \cite{Gunaydin:1985tc}.
\subsubsection{Joseph ideal of $SO(3,2)$ in Lorentz covariant basis}

To get a more physical picture of what the vanishing of the Joseph ideal means we shall go to  the conformal 3-graded basis defined in equation \ref{3dconfgrading}. Evaluating the Joseph ideal in this basis, we find that the vanishing ideal is equivalent to the linear combinations of certain quadratic identities, full set of  which hold only in the singleton realization. First we have the masslessness conditions:
\be
P^2=P^\mu P_\mu  =0 \quad, \quad  K^2 = K^\mu K_\mu=0
\ee
The remaining set of quadratic relations that define the Joseph ideal are
\bea
6\Delta \cdot \Delta +  2 M^{\mu\nu} \cdot M_{\mu\nu} + P^\mu \cdot K_\mu \eq 0 \\
P^\mu \cdot (M_{\mu\nu} + \eta_{\mu\nu} \Delta) \eq 0 \\
K^\mu \cdot (M_{\nu\mu} + \eta_{\nu\mu} \Delta) \eq 0 \\
 \eta^{\mu\nu} M_{\mu\rho} \cdot M_{\nu\sigma} -  P_{(\rho} \cdot K_{\sigma)}  +  \eta_{\rho\sigma} \eq 0 \\
\Delta \cdot M_{\mu\nu} + P_{[\mu} \cdot K_{\nu]} \eq 0 \\
M_{[\mu\nu} \cdot P_{\rho]} \eq 0 \\  M_{[\mu\nu} \cdot K_{\rho]} \eq 0
\eea
The ideal generated by these relations is completely equivalent to equation \ref{jos3d} but it sheds light on the massless nature of these representations. The scalar and spinor singleton modules for $SO(3,2)$ are the only minreps and there are no other deformations. This is a general phenomenon for all symplectic groups $Sp(2n,\mathbb{R})$ ($n>2$) and within the quasiconformal approach it can be explained by the fact that  the corresponding quartic invariant that enters the minimal unitary realization vanishes for symplectic groups.

The Casimir invariants for the singleton or the minrep of $SO(3,2)$ are as follows:
\bea
C_2 \eq M^{~A}_B M^{~B}_A = \frac{5}{2}  \\
C_4 \eq M^{~A}_B M^{~B}_C M^{~C}_D M^{~D}_A = -\frac{35}{8}
\eea
Computing the products of the generators of the above singleton realization corresponding to the Young tableaux $\parbox{10pt}{\YoungAAAA}$ and $\parbox{20pt}{\YoungB}$ one finds that  they vanish identically and the resulting enveloping algebra contains only the operators whose Young tableaux have two rows.
\subsection{Joseph ideal of  $SO(4,2)$ }
In this subsection we shall first evaluate the generator $J_{ABCD}$ of the Joseph ideal of $SO(4,2)$ in the covariant twistorial operator realization and then in the quasiconformal realization of its minrep to highlight the essential differences.
\subsubsection{Joseph ideal in the covariant twistorial oscillator or doubleton realization}
\label{jos4dtwist}
We will use the doubleton realization\cite{Gunaydin:1984fk,Gunaydin:1998sw}  reviewed in section \ref{4d-doubleton-section} to compute the generator $J_{ABCD}$  of the Joseph ideal for $SO(4,2)$ :
\bea
J_{ABCD} &=& M_{AB}M_{CD} - M_{AB}\circledcirc M_{CD} - \frac{1}{2}\comm{M_{AB}}{M_{CD}}
- \frac{1}{60} \langle M_{AB},M_{CD} \rangle \nn
&=& \frac{1}{2} M_{AB} \cdot M_{CD} - M_{AB}\circledcirc M_{CD}
- \frac{1}{60} \langle M_{AB},M_{CD} \rangle
\eea
Substituting the expressions for the generators in the covariant twistorial realization one finds that it does not vanish identically as an operator in contrast to the situation with the singletonic realization of $SO(3,2)$. However one finds that $J_{ABCD}$ has only 15 independent non-vanishing components which turn out to be equal to one of the following expressions  (up to an overall sign):
\bea
\lb (a_1 b_2 + a_2 b_1) \pm (a_2^\dagger b_1^\dagger + a_1^\dagger b_2^\dagger)
\rb\mathcal{Z}  && \\
\lb (a_1 b_2 - a_2 b_1) \pm (a_2^\dagger b_1^\dagger - a_1^\dagger b_2^\dagger)
\rb\mathcal{Z}  && \\
\lb (a_1 b_1 + a_2 b_2) \pm (a_2^\dagger b_2^\dagger + a_1^\dagger b_1^\dagger)\rb\mathcal{Z}
&&\\
\lb (a_1 b_1 - a_2 b_2) \pm (a_2^\dagger b_2^\dagger - a_1^\dagger b_1^\dagger)\rb\mathcal{Z}
&&\\
\lb a_1^\dagger a_2 + a_1a_2^\dagger \pm b_1^\dagger b_2 \pm b_1b_2^\dagger
\rb \mathcal{Z}  &&\\
\lb a_1^\dagger a_2 - a_1a_2^\dagger \pm b_1^\dagger b_2 \mp b_1b_2^\dagger
\rb \mathcal{Z}  &&\\
\lb (N_{a_1}-N_{a_2}) \pm (N_{b_1}-N_{b_2})\rb \mathcal{Z}  &&\\
(N_a+N_b+2)\mathcal{Z}  &&
\eea
Similarly, computing the products of the generators corresponding to the  the Young tableaux $\parbox{10pt}{\YoungAAAA}$ explicitly in the covariant twistorial realization one finds that they do not vanish  but we have the following relation:
\be
\parbox{10pt}{\YoungAAAA} = \mathcal{Z} \,\,\parbox{20pt}{\YoungAA}
\ee
Thus all non-vanishing four-row diagrams can be dualized to two-row diagrams and the resulting generators of the universal enveloping algebra in doubleton realization are described by two-row diagrams.

The operator $\mathcal{Z}= (N_a - N_b)$ commutes with all the generators of $SU(2,2)$ and its eigenvalues label the helicity of the corresponding massless representation of the conformal group \cite{Gunaydin:1984fk,Gunaydin:1998sw,Fernando:2009fq}.   All the components of the generator $J_{ABCD}$ of Joseph ideal vanish on the states that form the basis of an UIR of $SU(2,2)$ whose  lowest weight vector is the Fock vacuum $|0\rangle$ since
\be \mathcal{Z} |0\rangle =0 \ee
 The corresponding unitary representation  describes a conformal scalar field in four dimensions (zero helicity)  and is the true minrep of $SU(2,2)$ annihilated by  the Joseph ideal \cite{Fernando:2009fq}.

The Casimir invariants for $SO(4,2)$ in the doubleton representation are as follows:
\bea
\label{cas1}
C_2 \eq M^{~A}_B M^{~B}_A = \frac{3}{2} \lb 4-\mathcal{Z}^2\rb \\
C'_3 \eq  \epsilon^{ABCDEF}M_{AB}M_{CD}M_{EF} = 4 \mathcal{Z} C_2 = 8C_2 \sqrt{1-\frac{C_2}{6}} \\
C_4 \eq M^{~A}_B M^{~B}_C M^{~C}_D M^{~D}_A = \frac{C_2^2}{6} - 4 C_2
\label{cas2}
\eea
Thus we see that all the higher order Casimir invariants are functions of the quadratic Casimir $C_2$ which itself is given in terms of $\mathcal{Z}=N_a-N_b$.

\subsubsection{\label{josephqc4d} Joseph ideal and  the quasiconformal realization of the  minrep of $SO(4,2)$}
To apply Eastwood's formula to the generator of Joseph ideal in the quasiconformal realization  it is convenient to go from the conformal 3-graded basis to the $SO(4,2)$ covariant canonical basis where the generators $M_{AB}$ satisfy the following commutation relations:
\be
\comm{M_{AB}}{M_{CD}} = i (\eta_{BC}M_{AD}-\eta_{AC}M_{BD}-\eta_{BD}M_{AC}+\eta_{AD}M_{BC})
\ee
where the metric $\eta_{AB} = \text{diag}(-,+,+,+,+,-)$ is used to raise and lower the indices $A,B = 0,1,\ldots,5$ etc. In addition to Lorentz generators $M_{\mu\nu}$ ($\mu, \nu, ..=0,1,2,3$) , we have  the following linear relations between the generators in the canonical basis and the  conformal 3-graded basis
\bea
M_{\mu 4} \eq \frac{1}{2} \lb P_\mu - K_\mu \rb \\
M_{\mu 5} \eq \frac{1}{2} \lb P_\mu + K_\mu \rb \\
M_{45} \eq -\Delta
\eea
Substituting the  expressions for the quasiconformal realization of  the generators of $SO(4,2)$ given in subsection \ref{qcminrep}  into the generator of the Joseph ideal in the canonical basis:
\be
J_{ABCD} = \frac{1}{2} M_{AB}\cdot M_{CD} - M_{AB}\circledcirc M_{CD} - \frac{1}{60} \langle M_{AB},M_{CD} \rangle
\label{4d-joseph}
\ee
one finds that it {\it vanishes identically as an  operator}  showing that the corresponding unitary representation is indeed the minimal unitary representation.

We should stress the important point that the tensor product of the Fock spaces of the two oscillators $d$ and $g$ with the state space of the singular oscillator $A_{\mathcal{L}} \, (A^\dagger_{\mathcal{L}})$ form the basis of a single UIR which is  the minrep. In contrast, the Fock space of the covariant twistorial oscillators reviewed  in section \ref{4d-doubleton-section}   decomposes into infinitely  many UIRs (doubletons)  of which only the irreducible representation whose lowest weight vector is the Fock vacuum, which is annihilated by $J_{ABCD}$,  is the minimal unitary representation of $SO(4,2)$.


\subsubsection{$4d$ Lorentz Covariant Formulation of the  Joseph ideal of $SO(4,2)$}
Above we showed that the  generator of the Joseph ideal given in equation (\ref{4d-joseph})  vanishes identically
as an operator for the quasiconformal realization of the minrep of $SO(4,2)$ in the canonical basis.   To get a more physical picture of what the vanishing of  generator $J_{ABCD}$  means we shall go to  the conformal three grading  defined by the dilatation generator $\Delta$. Evaluating the generator of the Joseph ideal in this basis, we find that the vanishing condition is equivalent to  linear combinations of certain quadratic identities, full set of  which hold only in the quasiconformal realization of the minrep. First we have the conditions:
\be
P^2=P^\mu P_\mu  =0 \quad, \quad  K^2 = K^\mu K_\mu=0
\ee
which hold also for the twistorial oscillator realization given in section \ref{4d-doubleton-section}.
The remaining set of quadratic relations that define the Joseph ideal are
\bea
4\Delta \cdot \Delta +  M^{\mu\nu} \cdot M_{\mu\nu} + P^\mu \cdot K_\mu \eq 0 \\
P^\mu \cdot (M_{\mu\nu} + \eta_{\mu\nu} \Delta) \eq 0 \\
K^\mu \cdot (M_{\nu\mu} + \eta_{\nu\mu} \Delta) \eq 0
\eea
\bea
 \eta^{\mu\nu} M_{\mu\rho} \cdot M_{\nu\sigma} -  P_{(\rho} \cdot K_{\sigma)} + 2 \eta_{\rho\sigma} \eq 0 \label{mpident}\\
M_{\mu\nu} \cdot M_{\rho\sigma} + M_{\mu\sigma}\cdot M_{\nu\rho} + M_{\mu\rho}\cdot M_{\sigma\nu} \eq 0  \label{mident}\\
\Delta \cdot M_{\mu\nu} + P_{[\mu} \cdot K_{\nu]} \eq 0 \label{dmident}
\eea
\be
M_{[\mu\nu} \cdot P_{\rho]} = 0, \qquad \qquad M_{[\mu\nu} \cdot K_{\rho]} = 0
\label{helicity}
\ee
In four dimensions, using the Levi-Civita tensor one can define the Pauli-Lubanski vector, $W^\mu$ and its conformal analogue, $V^\mu$ as follows:
\be
W^\mu = \frac{1}{2}\epsilon^{\mu\nu\rho\sigma}P_\nu M_{\rho\sigma}, \qquad V^\mu = \frac{1}{2}\epsilon^{\mu\nu\rho\sigma}K_\nu M_{\rho\sigma}
\ee
where $\epsilon_{0123}=+1, \epsilon^{0123}=-1$ and the indices are raised and lowered by the Minkowski metric.
For massless fields, $W^\mu$ and $V^\mu$ are proportional to $P^\mu$ and $K^\mu$ respectively with the proportionality constant related to helicity of the fields \cite{weinberg1996quantum,Siegel:1999ew}. Equations (\ref{helicity}) imply that for the minrep both the $W^\mu$ and $V^\mu$ vanish implying that it describes a zero helicity (scalar) massless  field \footnote{ We should note that a similar set of identities (constraints) were discussed in \cite{Siegel:1999ew} in the context of deriving field equations for particles of all spins (acting on field strengths) where they arise as conformally covariant forms  of massless particles. }.

Computing the products of the generators in the above quasiconformal realization corresponding to Young tableaux $\parbox{10pt}{\YoungAAAA}$ and $\parbox{20pt}{\YoungB}$ explicitly one finds  that they vanish identically and the resulting enveloping algebra contains only operators with two row Young tableaux.

The Casimir operators of  the minrep of $SO(4,2)$ are as take on the following values (using the definitions given in equations \ref{cas1} - \ref{cas2}):
\be
C_2 = 6, \qquad C_3 = 0 \qquad C_4 = -18.
\ee


\subsection{Deformations of the minrep of $SO(4,2)$ and their  associated  ideals}
\label{sect-def-joseph}
As was shown in \cite{Fernando:2009fq}, the minimal unitary representation of $SU(2,2)$  that corresponds to a conformal scalar field admits a one-parameter ($\zeta$) family of deformations corresponding to massless conformal fields of helicity $\frac{\zeta}{2}$ in four dimensions, which can be continuous. For non-integer values of the deformation parameter $\zeta$ they correspond, in general, to unitary representations of an infinite covering of the conformal group\footnote{Recently, such a  continuous helicity parameter was introduced as a spectral parameter for scattering amplitudes in N=4 super Yang-Mills theory in \cite{Ferro:2012xw}.}.

The generators of the deformed minreps were reformulated in terms of deformed twistorial oscillators in section \ref{sect4dzeta}. Substituting  the expressions for the generators of the deformed minreps of $SO(4,2)$ into the generator   $J_{ABCD}$ of the Joseph ideal one finds that it  does not vanish for non-zero values of the deformation parameter $\zeta$. One might therefore ask  if there exists deformations of the Joseph ideal that annihilate the deformed minimal unitary representations labelled by the deformation parameter $\zeta$.  Remarkably, this is indeed the case.   The quadratic identities that define the Joseph ideal in the conformal basis discussed in the previous section go over to identities involving the  deformation parameter $\zeta$ and define the deformations of the Joseph ideal.
One finds that the helicity conditions are modified as follows:
\bea
\half \epsilon^{\mu\nu\rho\sigma}M_{\nu\rho} \cdot P_{\sigma} \eq \zeta P^\mu \\
\half \epsilon^{\mu\nu\rho\sigma}M_{\nu\rho} \cdot K_{\sigma} \eq -\zeta K^\mu
\eea
The identities (\ref{mpident}), (\ref{mident}) and (\ref{dmident}) get also modified as follows:
\bea
 \eta^{\mu\nu} M_{\mu\rho} \cdot M_{\nu\sigma} -  P_{(\rho} \cdot K_{\sigma)} + 2 \eta_{\rho\sigma} \eq \frac{\zeta^2}{2} \eta_{\rho\sigma} \\
M_{\mu\nu} \cdot M_{\rho\sigma} + M_{\mu\sigma}\cdot M_{\nu\rho} + M_{\mu\rho}\cdot M_{\sigma\nu}   \eq  \zeta\epsilon_{\mu\nu\rho\sigma} \Delta \\
\Delta \cdot M_{\mu\nu} + P_{[\mu} \cdot K_{\nu]} \eq  - \frac{\mathcal{\zeta}}{2} \epsilon_{\mu\nu \rho\sigma}M^{\rho\sigma}
\eea
The other  quadratic identities remain unchanged in going over to the deformed minimal unitary representations .

The Casimir invariants for the deformations of the minrep of $SO(4,2)$ depend only the deformation parameter $\zeta$  and are given as follows (using the definitions given in equations \ref{cas1} - \ref{cas2}):
\bea
C_2 \eq 6 - \frac{3\zeta^2}{2} \\
C'_3 \eq 6 \zeta \lb \zeta^2-4 \rb = -8C_2 \sqrt{1-\frac{C_2}{6}} \\
C_4 \eq \frac{3}{8} \lb \zeta^4 + 8\zeta^2 -48 \rb = \frac{C_2^2}{6} - 4 C_2
\eea

The products of generators corresponding to the Young tableaux $\parbox{10pt}{\YoungAAAA}$  do not vanish for the deformed minimal unitary representations and depend on the deformation parameter $\zeta$ as follows:

\be
\parbox{10pt}{\YoungAAAA} = \zeta \,\,\parbox{20pt}{\YoungAA}
\ee
Hence all non-vanishing four-row diagrams can be dualized to two-row diagrams and the resulting generators of the universal enveloping algebra in deformed realization are described by two-row diagrams.  For $\zeta=0 $ all four row diagrams  vanish and  one obtains the standard high spin algebra of Vasiliev type whose generators transform in representations of the underlying $AdS$ group corresponding to Young tableaux containing only two row traceless diagrams. For non-zero $\zeta$ the corresponding enveloping algebras describe deformed higher spin  algebras as discussed in the next subsection.

We saw earlier in section \ref{jos4dtwist} that the generator $J_{ABCD}$  of the Joseph ideal  did not vanish identically as an operator for the covariant twistorial oscillator realization of $SO(4,2)$. It annihilates only the states belonging to  the subspace that form the basis of the true minrep of $SO(4,2)$.  By going to the conformal three grading, one finds that the generator $J_{ABCD}$ of the Joseph ideal can be written in a form similar to the deformed quadratic identities above with the deformation parameter replaced by the linear Casimir operator $\mathcal{Z}=N_a-N_b$
\bea
 \eta^{\mu\nu} M_{\mu\rho} \cdot M_{\nu\sigma} - P_{(\rho} \cdot K_{\sigma)}  + 2 \eta_{\rho\sigma} \eq \frac{\mathcal{Z}^2}{2} \eta_{\rho\sigma} \\
M_{\mu\nu} \cdot M_{\rho\sigma} + M_{\mu\sigma}\cdot M_{\nu\rho} + M_{\mu\rho}\cdot M_{\sigma\nu}   \eq -\mathcal{Z}\epsilon_{\mu\nu\rho\sigma} \Delta \\
\Delta \cdot M_{\mu\nu} + P_{[\mu} \cdot K_{\nu]} \eq  \frac{\mathcal{Z}}{2} \epsilon_{\mu\nu \rho\sigma}M^{\rho\sigma}\\
\half \epsilon^{\mu\nu\rho\sigma}M_{\nu\rho} \cdot P_{\sigma} \eq -\mathcal{Z} P^\mu \\
\half \epsilon^{\mu\nu\rho\sigma}M_{\nu\rho} \cdot K_{\sigma} \eq \mathcal{Z} K^\mu
\eea
The Fock space of the oscillators decompose into an infinite set of unitary irreducible representations of $SU(2,2)$ corresponding to massless conformal fields of all integer and  half-integer helicities labelled by the eigenvalues of $\mathcal{Z}/2$.


\subsection{Higher spin algebras and superalgebras and their deformations}
\label{hssect}
We shall adopt the definition of the higher spin $AdS_{(d+1)}/CFT_d$ algebra as the quotient of the enveloping algebra  $\mathscr{U}(SO(d,2))$  of $SO(d,2)$ by the Joseph ideal $\mathscr{J}(SO(d,2))$ and denote it as $HS(d,2) $ \cite{Eastwood:2002su}:
\be
HS(d,2) = \frac{\mathscr{U}(SO(d,2))}{\mathscr{J}(SO(d,2))}
\ee
We shall however extend it to define deformed higher spin algebras as the enveloping algebras  of the deformations of the  minreps of the corresponding $AdS_{d+1}/Conf_d$ algebras. For these deformed higher spin algebras  the corresponding deformations of the
Joseph ideal vanish identically as operators in the quasiconformal realization as we showed explicitly above
for the conformal group in four dimensions.
 We expect a given deformed higher spin algebra to be  the {\it unique } infinite dimensional quotient of the universal enveloping algebra of  an appropriate covering\footnote{In $d=4$ deformed minreps describe massless fields with the helicity $\frac{\zeta}{2}$. For non-integer values of $\zeta$ one has to go to an infinite covering of the $4d$ conformal group.} of the conformal group  by the deformed ideal as was shown  for the undeformed minrep in \cite{MR2369839}.   Similarly, we define the higher spin  superalgebras and their deformations as the enveloping algebras of the minimal unitary realizations of the underlying superalgebras and their deformations, respectively\footnote{We should note that  the universal enveloping algebra of a Lie group as defined in the mathematics literature is an associative algebra with unit element. Under the commutator product inherited from the underlying Lie algebra it becomes a  Lie algebra.}.

In four dimensions $(d=4)$ we have a one parameter family of higher spin algebras  labelled by the helicity $\zeta/2$:
\be
HS(4,2;\zeta) = \frac{\mathscr{U}(SO(4,2))}{\mathscr{J}_\zeta(SO(4,2))} \label{bosonicalgebras}
\ee
where $\mathscr{J}_\zeta(SO(4,2))$ denotes the deformed Joseph ideal of $SO(4,2)$ defined in section \ref{sect-def-joseph}
\footnote{After the main results of this paper was announced at the GGI Conference
on higher spin theories in May 2013, we became aware of the work of  \cite{Boulanger:2011se} where possible
deformations of purely bosonic higher spin algebras in arbitrary dimensions
were studied and it was shown that the deformations can depend at most
on one parameter. In a subsequent work \cite{Boulanger:2013zza}  it was shown that the $AdS_d$ higher spin algebra is unique in $d = 4$ and $d > 6$ under their assumptions on the spectrum of
generators. They also imply that the one parameter family of deformations of  \cite{Boulanger:2011se} must be the same as the one parameter family discussed in this paper which is based on earlier work \cite{Fernando:2009fq}  that they cite. The results of \cite{ Boulanger:2011se} are based on Young tableaux analysis of gauge fields in $AdS_5$.
Whether and how their deformation parameter is related to helicity in $4d$ is
not known at this point.
Furthermore, we find a discrete infinite family of higher spin algebras and
superalgebras in $AdS_7$\cite{Govil:2014uwa}.}

 On the $AdS_{d+1}$ side the generators of the higher spin algebras correspond to higher spin gauge fields, while on the $Conf_d$  side they are related to conserved tensors including the conserved stress-energy tensor. The charges associated with the generators of conformal algebra $SO(d,2)$ are defined by conserved currents constructed by contracting the stress-energy tensor with conformal Killing vectors. Similarly, the higher conserved currents are obtained by contracting the conformal Killing tensors with the stress-energy tensor. These higher conformal Killing tensors are obtained simply by tensoring the conformal Killing vectors with themselves. Though implicit in previous work on the subject \cite{Konstein:2000bi,Vasiliev:2001wa}, explicit use of the language of conformal Killing tensors in describing  higher spin algebras seems to have first appeared in the paper of Mikhailov \cite{Mikhailov:2002bp}\footnote{The  generators of the higher spin algebra $shs^E(8|4)$ of reference \cite{Konshtein:1988yg} in terms of  Killing spinors in $AdS_4$ were obtained in the work of \cite{Sezgin:1998gg} on higher spin $N=8$ supergravity in $d=4$. }. This connection was put on a rigorous foundation by Eastwood in his study of the higher symmetries of the Laplacian \cite{Eastwood:2002su} who showed that the undeformed higher spin algebra  can be obtained as the quotient of the enveloping algebra of $SO(d,2)$ generated by the conformal Killing tensors quotiented by the Joseph ideal.

The connection between the doubleton realization of $SO(4,2)$ in terms of covariant  twistorial oscillators and the corresponding conformal Killing tensors was also studied by Mikhailov \cite{Mikhailov:2002bp}. He pointed out that the higher conformal Killing tensors  correspond to the products of bilinears of oscillators that generate $SO(4,2)$  in the doubleton realization of \cite{Gunaydin:1984fk,Gunaydin:1998sw,Gunaydin:1998jc}, which are elements of the enveloping algebra.  The undeformed higher spin algebra $HS(4,2;\zeta=0)$ was also studied by the authors of \cite{Sezgin:2001zs} who  used  the doubletonic construction as well\footnote{ We should note that the infinite dimensional ideal that appears  in the work of \cite{Sezgin:2001zs} is not the Joseph ideal. }. The undeformed higher spin algebra  corresponding to  \ref{bosonicalgebras} describes fields of all spins  which are multiplicity-free. The model studied in detail in \cite{Sezgin:2001zs} is based on the  minimal infinite dimensional subalgebra of $HS(4,2;\zeta=0)$ that describes  only even spins. For the purely bosonic higher spin algebras in $AdS_d$ the factorization of the ideal generated by the bilinear operators corresponding to the last three irreps in \ref{ytdecomp} was discussed in great detail, at the level of abstract enveloping algebra, in \cite{Iazeolla:2008ix} without reference to explicit oscillator realization. They corroborated that in $d=5$ there is an equivalence between factoring out this ideal and the oscillator construction in  \cite{Sezgin:2001zs}.

The supersymmetric extension of the  higher spin algebras $HS(4,2;\zeta) $ is given by the enveloping algebra of the deformed minimal unitary realization of the N-extended conformal superalgebras $SU(2,2|N)_\zeta$ with the even subalgebras $SU(2,2)\oplus U(N)$. We shall denote the resulting higher spin algebra as $HS[SU(2,2|N);\zeta]$.  These  supersymmetric extensions  involve odd powers of the deformed twistorial oscillators and  the identities that define the Joseph ideal get extended to a supermultiplet of identities obtained by the repeated actions of $Q$ and $S$ supersymmetry generators on the generators of Joseph ideal.  On the conformal side the odd generators  correspond to the products of  the conformal Killing spinors with conformal Killing vectors and  tensors.
If we denote the resulting deformed super-ideal as $\mathscr{J}_{\zeta}[SU(2,2|N)]$ we can formally write
\be
HS[SU(2,2|N);\zeta]= \frac{\mathscr{U}(SU(2,2|N)}{\mathscr{J}_{\zeta}[SU(2,2|N)]}
\ee

As discussed in previous sections, the bosonic higher spin gauge fields are described by two-row Young tableaux of $SO(4,2)$.  However in order to extend the $SO(4,2)$ Young tableaux to super Young tableaux , one needs to represent the corresponding representations  in terms of  $SU(2,2)$ Young tableaux  since the relevant superconformal algebras are $SU(2,2|N)$ with the even subalgebra $SU(2,2) \oplus U(N)$ and not  superalgebras of the orthosymplectic type. The  identifications of Young tableaux can be easily checked by calculating the dimensions of corresponding irreps. In going from  $SU(2,2)$ to $SU(2,2|N)$ one simply replaces the Young tableaux  of $SU(2,2)$ by super Young tableaux of $SU(2,2|N)$\footnote{For a review of super Young tableaux we refer to \cite{Bars:1982ps} and a study of the relations  between Super young tableaux and Kac-Dynkin labelling see \cite{Bars:1982se}.}.   The Young diagram of the adjoint representation of $SO(4,2)=SU(2,2)$ goes over to the following  super tableau of $SU(2,2|N) $ which involves one dotted and one undotted ``superboxes" :
\be
\overbrace{\oneonebox}^{SO(4,2)} \Longleftrightarrow  \overbrace{\twooneonebox}^{SU(2,2)} \xrightarrow{\text{supersymmetrize}} \overbrace{\stwooneonebox}^{SU(2,2|N)}
\ee
Notice that the $\oneoneonebox$ of $SU(2,2)$ is represented as $\donebox$.  This is because supersymmetric extensions  of $SU(2,2)$ do not have an invariant super Levi-Civita tensor. The adjoint representation $\stwooneonebox$ os $SU(2,2|N)$ can be decomposed with respect to $SU(2,2)\times U(N)$ as follows:
\be
\stwooneonebox = ( \dtwooneonebox, 1 ) \oplus ( \donebox, \onebox ) \oplus ( \onebox, \donebox ) \oplus ( 1, \dtwooneonebox ) \oplus ( 1,1 )
\ee

The window diagram appearing in the tensor product of two adjoint representations can be supersymmetrized as follows:
\be
\overbrace{\twotwobox}^{SO(4,2)} \Longleftrightarrow  \overbrace{\ttwooneonebox}^{SU(2,2)} \xrightarrow{\text{supersymmetrize}} \overbrace{\sttwooneonebox}^{SU(2,2|N)}
\ee
Thus the higher spin gauge fields are described by the following $SU(2,2)$ diagram and can be consequently supersymmetrized in a straightforward manner:
\be
\marctwojtwojbox \Longleftrightarrow  \overbrace{\tmarctwojtwojbox}^{SU(2,2)} \xrightarrow{\text{supersymmetrize}} \overbrace{\stmarctwojtwojbox}^{SU(2,2|N)}
\ee

Thus the generators of the supersymmetric extension of the undeformed ($\zeta =0$) bosonic higher spin algebra decompose as
\be
HS[SU(2,2|N);0] =\sum_\oplus  \overbrace{\stmarctwojtwojbox}^{SU(2,2|N)}
\ee
For the decomposition of the deformed higher spin algebras for integer values of the deformation parameter one can also use the super tableaux to represent their generators. However for non-integer values of the deformation parameter it may be necessary to use Kac-Dynkin or other labellings.
We should stress the important point that the maximal finite dimensional subalgebra of $HS[SU(2,2|N);\zeta]$ is the superconformal algebra $SU(2,2|N)_\zeta$ with $\zeta$ labelling the central charge. The decomposition of the generators of the higher spin superalgebra with respect to the $SU(2,2|N)_\zeta$ subalgebra is multiplicity free.

Furthermore, the minimal unitary realization of the supersymmetric extensions $SU(2,2|N)$ of $SU(2,2)$ the enveloping algebra of $SU(2,2|N)$ has as subalgebras the enveloping algebras of different unitary representations of $SU(2,2)$ that form the minimal unitary supermultiplet modded out by the corresponding deformed Joseph ideals. These different irreps of $SU(2,2)$ are deformations of the minrep where the deformation is driven by the fermionic oscillators and the deformation parameter $\zeta$ is given by the eigenvalue of the number operator of these fermionic oscillators $ N_\xi $ in the minimal unitary supermultiplet. In the deformations of the minimal unitary supermultiplet labelled by $\zeta$ the enveloping algebra of $SU(2,2|N)$ has a similar decomposition in which the irreps making of the supermultiplet are labelled by a linear combination of $\zeta$ and the eigenvalue of $N_\xi$.

 We should note that the undeformed super algebra $HS[SU(2,2|4);\zeta=0]$ for the particular value of $N=4$ was studied in \cite{Sezgin:2001yf} by using the doubletonic realization. As in the purely bosonic case  the higher spin theory studied in  \cite{Sezgin:2001yf} is based on a subalgebra of $HS[SU(2,2|4);\zeta=0]$ and  the resulting linearized gauging was shown to correspond to massless $PSU(2,2|4)$ multiplets whose maximal spins are even integers $2,4,...$.\footnote{ We should stress again that   the infinite dimensional ideal that appears in the work of \cite{Sezgin:2001yf} is not the Joseph ideal.}

The situation is much simpler for $AdS_4/Conf_3$ higher spin algebras.  There are only two minreps of $SO(3,2)$, namely the scalar and spinor singletons.  The higher spin algebra $HS(3,2)$ is simply given by the enveloping algebra of the singletonic realization of $Sp(4,\mathbb{R})$ \cite{Gunaydin:1989um,Konshtein:1988yg}. Singletonic realization  of the Lie algebra of $Sp(4,\mathbb{R})$ describes both the scalar and spinor singletons. They form  a single irreducible supermultiplet of $OSp(1|4,\mathbb{R})$ generated by taking the twistorial oscillators as the odd generators \cite{ Gunaydin:1989um}. The odd generators  correspond to  conformal Killing spinors , which together with conformal Killing vectors in $d=3$ , generate the Lie super-algebra of $OSp(1|4,\mathbb{R})$.  Its enveloping algebra leads to the higher spin super-algebra of Fradkin-Vasiliev type  involving  all integer and half integer spin fields in $AdS_4$. One can also construct N-extended higher spin superalgebras in $AdS_4$ as enveloping algebras of the singletonic realization of $OSp(N|4,\mathbb{R})$ \cite{Gunaydin:1989um,Konshtein:1988yg}.


\section{Discussion}
\label{section-discussion}
The existence of a one-parameter family of  $AdS_5/Conf_4$ higher spin algebras and superalgebras raises the question as to their physical meaning.  Before discussing the situation in four dimensions, let us summarize what is known for $AdS_4/Conf_3$ higher spin algebras. In $3d$, there is no deformation of the higher spin algebra except for the super extension corresponding to $Sp(4;\mathbb{R}) \rightarrow OSp(N|4;\mathbb{R})$. For the bosonic $AdS_4$ higher spin algebras one finds that the higher spin theories of Vasiliev are dual to certain conformally invariant vector-scalar/spinor models in $3d$ \cite{Klebanov:2002ja,Sezgin:2003pt} in the large N limit.
More recently, Maldacena and Zhiboedov studied the constraints imposed on a conformal field theory in $d=3$ three dimensions by the existence of a single conserved  higher spin current. They found that this implies the existence of an infinite number of conserved higher spin currents. This corresponds simply to the fact that the generators of $SO(d, 2)$ get extended to an infinite spin algebra when one takes their commutators with an operator which is  bilinear or higher order in the generators, except for those operators that correspond to the Casimir elements. They also showed that the correlation functions of the stress tensor and the conserved currents are those of a free field theory in three dimensions, either a theory of $N$ free bosons or a theory of $N$ free fermions \cite{Maldacena:2011jn}, which are simply the scalar and spinor singletons.

The distinguishing feature of $3d$ is the fact that there exists only two minimal unitary representations corresponding to massless conformal fields  which are simply the  $Di$ and $Rac$ representations of Dirac. However in $4d$, we have a one-parameter, $\zeta$, family of deformations of the minimal unitary representation of the conformal group corresponding to massless conformal fields of  helicity $\zeta/2$. The same holds true for the minimal unitary supermultiplet of $SU(2,2|N)$. From M/superstring point of view the most important interacting and supersymmetric CFT in $d=4$ is the $N=4$ super Yang-Mills theory.
It was argued in references  \cite{Sundborg:2000wp,Sezgin:2001zs,witten2001spacetime} that the holographic dual of $\mathcal{N}=4$ super Yang-Mills theory with gauge group $SU(N)$ at $g^2_{YM}N=0$ for $N \rightarrow \infty$ should be a free gauge invariant theory in $AdS_5$ with massless fields of arbitrarily high spin and this  was supported by calculations in \cite{Mikhailov:2002bp}. Moreover the scalar sector of $\mathcal{N}=4$ super Yang-Mills theory at $g^2_{YM}N=0$ with $N \rightarrow \infty$ should be dual to bosonic higher spin theories in $AdS_5$ which provides a non-trivial extension of $AdS/CFT$ correspondence in superstring theory \cite{Maldacena:1997re,Witten:1998qj} to non-supersymmetric large $N$ field theories.
The Kaluza-Klein spectrum of IIB supergravity over $AdS_5\times S^5$ was first obtained by tensoring of the minimal unitary supermultiplet (scalar doubleton) of $SU(2,2|4)$   with itself repeatedly and restricting to CPT self-conjugate
sector \cite{Gunaydin:1984fk}. The massless graviton supermultiplet in $AdS_5$ sits at the bottom of this infinite tower. In fact all the unitary representations corresponding to massless fields in $AdS_5$ can be obtained by tensoring of two doubleton representations of $SU(2,2)$ which describe massless conformal fields on the boundary of $AdS_5$ \cite{Gunaydin:1984fk,Gunaydin:1989um,Gunaydin:1998sw,Gunaydin:1998jc}.
As was argued by Mikhailov \cite{Mikhailov:2002bp}, in the large N limit the correlation functions in the CFT side become products of two point functions which correspond to products of two doubletons. As such  they correspond to massless fields in the $AdS_5$ bulk. At the level of correlation functions the same arguments suggest  that corresponding to  a one parameter family of deformations of the N=4 Yang-Mills supermultiplet there must exists a family of  supersymmetric massless higher spin theories in $AdS_5$.  Turning on the gauge coupling constant on the Yang-Mills side leads to interactions in the bulk and most of the higher spin fields  become massive.

The fact that the quasiconformal realization of the minrep of $SU(2,2|4)$ is nonlinear implies that the corresponding higher spin theory in the bulk must be interacting.
Since the same minrep can also be obtained by using the doubletonic realization \cite{Gunaydin:1984fk}, which corresponds to free field realization, suggests that the interacting supersymmetric higher spin theory may be integrable in the sense that its holographic dual is an integrable conformal  field theory with infinitely many conserved currents , just like the classical $N=4$ super Yang-Mills theory in four dimensions \cite{Bena:2003wd}.
There are other deformed higher spin algebras corresponding to non-CPT self-conjugate
supermultiplets of $SU(2,2|4)$ that contain scalar fields and are deformations of the minrep.  The above arguments suggest that they too should  correspond to interacting but integrable supersymmetric higher spin theories in $AdS_5$. One solid piece of the evidence for this is provided by the fact that the symmetry superalgebras of interacting (nonlinear) superconformal quantum mechanical models of \cite{Fedoruk:2009xf} furnish a one parameter family of deformations of the minimal unitary representation of the $N=4$ superconformal algebra $D(2,1;\alpha)$ in one dimension. This was predicted in \cite{Gunaydin:2007vc} and shown explicitly  in \cite{Govil:2012rh}.

Most of the work on higher spin algebras until now have utilized the  realizations of underlying Lie (super)algebras as bilinears of oscillators which correspond to free field realizations. The quasiconformal approach allows one to give a natural definition of super Joseph ideal and leads directly to the interacting realizations of the superextensions of higher spin algebras.
The next step in this approach is to reformulate these interacting quasiconformal realizations in terms of covariant gauge fields and construct Vasiliev type nonlinear theories of interacting higher spins in $AdS_5$.

Another application of our results will be to reformulate the spin chain models associated with $N=4$ super Yang-Mills theory  in terms of deformed twistorial oscillators  and study  the  integrability  of corresponding spin chains non-perturbatively. In fact, a spectral parameter related to helicity and central charge was introduced recently for scattering amplitudes in $N=4$ super Yang-Mills theory \cite{Ferro:2013dga}. This spectral parameter corresponds to our deformation parameter which is helicity and appears as a central charge in the quasiconformal realization of the super algebra $SU(2,2|4)$.
We hope to address these issues in future investigations.

{\bf Acknowledgements:} Main results of this paper were announced in GGI Workshop on Higher spin symmetries (May 6-9, 2013)  and Summer Institute (Aug. 2013)  at ENS in Paris by MG and at Helmholtz Internations Summer School (Sept. 2013) at Dubna by KG. We would like to thank the organizers of these workshops and institutes  for their kind hospitality where part of this work was  carried out. We enjoyed discussions with many of their participants. We are especially grateful  to Misha Vasiliev and Eugene Skvortsov for stimulating discussions regarding higher spin theories and Dmytro Volin regarding quasiconformal realizations. This research was supported in part by the US National Science Foundation under grants PHY-1213183, PHY-08-55356 and DOE Grant No: DE-SC0010534.

\begin{appendices}

\section{Spinor conventions for $SO(2,1)$}
\label{so21}
We follow \cite{Samtleben:2009ts} for the spinor conventions in $d=3$ and thus all the $3d$ spinors are Majorana with $\eta_{\mu\nu} = \text{diag}(-,+,+)$. The gamma-matrices in Majorana representation terms of the Pauli matrices $\sigma^i$ are as follows:
\be
\gamma^0=-i\sigma^2, \quad \gamma^1=\sigma^1, \quad \gamma^2= \sigma^3
\ee
and they satisfy
\begin{equation}
\label{eq:a1}
\{\gamma^\mu,\gamma^\nu\}^\alpha_{\s\beta}=2\eta^{\mu\nu}\delta^\alpha{}_{\beta}\ \ .
\end{equation}
Thus the matrices $\lb \gamma^{\mu}\rb^\alpha_{\s \beta}$ are real and the Majorana condition on spinors imply
that they are real two component spinors. Spinor indices are raised/lowered by the epsilon symbols with
$\epsilon^{12}=\epsilon_{12}=1$ and choosing NW-SE conventions
\begin{equation}
\label{eq:a2}
\epsilon^{\alpha\gamma}\epsilon_{\beta\gamma}=\delta^\alpha_{\ \beta}\  ,\quad
\lambda^\alpha:=\epsilon^{\alpha\beta}\lambda_\beta\Leftrightarrow\lambda_\beta
    = \lambda^\alpha\epsilon_{\alpha\beta}\ .
\end{equation}

Introducing the real symmetric matrices
$\lb \sigma^\mu\rb_{\alpha\beta}:=\lb\gamma^\mu\rb^ \rho_{\s\beta}\  \epsilon_{\rho\alpha}$ and
$\lb\bar{\sigma}^{\mu}\rb^{\alpha\beta}:= (\epsilon \cdot \sigma^\mu \cdot \epsilon)^{\alpha\beta}
=-\epsilon^{\beta\rho}\  \lb\gamma^\mu\rb^\alpha_{\s\rho} $, a three vector in spinor notation writes as
a symmetric real matrix as
\begin{equation}
\label{eq:a3}
V_{\alpha\beta}:= \lb\sigma^\mu V_\mu \rb_{\alpha\beta} \ \ \Rightarrow \ \
 V^\mu=\frac{1}{2}\lb \bar{\sigma}^\mu \rb^{\alpha\beta}V_{\alpha\beta}\ .
\end{equation}
\section{Minimal unitary supermultiplet of $OSp(N|4,\mathbb{R})$ \label{OSp_N_4}}
In this section we will formulate the minimal unitary representation of $OSp(N|4,\mathbb{R})$ which is the superconformal algebra with $N$ supersymmetries in three dimensions. The superalgebra $\mathfrak{osp}(N|4)$ can be given a five graded decomposition with respect to the noncompact dilatation generator $\Delta$ as follows:
\be
\mathfrak{osp}(N|4) = K^{\a\b} \oplus S^{I\a} \oplus (O^I_J \oplus M_\a^{\s\b} \oplus \Delta) \oplus Q_\a^{\s I} \oplus P_{\a\b}
\ee
We shall call this superconformal 5-grading. The bosonic conformal generators are the same as given in previous section. In order to realize the $R$-symmetry algebra $SO(N)$  and supersymmetry generators, we introduce Euclidean Dirac gamma matrices  $\gamma^I$ ($I,J=1,2,\ldots,N$) which satisfy
\be
\acomm{\gamma^I}{\gamma^J} = \d^{IJ}
\ee
The $R$-symmetry generators are then simply given as follows:
\be
O^{IJ} = \frac{1}{2} \lb \gamma^I  \gamma^J  - \gamma^J  \gamma^I \rb
\ee
and supersymmetry generators are the bilinears of $3d$ twistorial oscillators and $\gamma^I$:
\be
Q_\a^{\s I} = \kappa_\a \gamma^I, \qquad S^{I\a} = \gamma^I \mu^\a
\ee
They satisfy the following commutation relations:
\be
\acomm{Q_\a^{\s I}}{Q_\b^{\s J}} = \d^{IJ}P_{\a\b}, \qquad \acomm{S^{I\a}}{S^{J\b}} = \d^{IJ}K^{\a\b}
\ee
\be
\acomm{Q_\a^{\s I}}{S^{J\b}} = M_\a^{\s\b} \d^{IJ} + 2 O^{IJ}\d_a^\b +i \lb \Delta - \frac{i}{2} \rb \d^{IJ}\d_\a^\b
\ee
The action of conformal group generators on supersymmetry generators is as follows:
\be
\comm{M_\a^{\s\b}}{Q_\g^{\s I}} = -\d_\g^\b Q_\a^{\s I} + \frac{1}{2}Q_\g^{\s I}, \qquad \comm{M_\a^{\s\b}}{S^{I\g}} = \d_\g^\a S^{I\b} - \frac{1}{2}S^{I\g}
\ee
\be
\comm{\Delta}{Q_\a^{\s I}} = \frac{i}{2}Q_\a^{\s I}, \qquad \comm{\Delta}{S^{I\a}} = -\frac{i}{2}S^{I\a}
\ee
\be
\comm{P_{\a\b}}{S^{I\g}} =-2 \d^\g_{(\a}Q_{\b)}^{\s I}, \qquad \comm{K^{\a\b}}{Q_\g^{\s I}} =2 \d_\g^{(\a}S^{\b) I}
\ee
\be
\comm{P_{\a\b}}{Q_\g^{\s I}} = \comm{K^{\a\b}}{S^{I\g}} = 0
\ee
The $R$-symmetry generators act only on the $I,J$ indices and rotate them as follows:
\be
\comm{Q_\a^{\s I}}{O^{JK}} = \frac{1}{2} \d^{I[J} Q_\a^{\s K]}, \qquad \comm{S^{I\a}}{O^{JK}} = \frac{1}{2} \d^{I[J} S^{K]\a}
\ee

For even $N$ the  singleton supermultiplet consists of conformal scalars transforming in a chiral spinor representation of $SO(N)$ and conformal space-time spinors transforming in the opposite chirality spinor representation of $SO(N)$ \cite{Gunaydin:1985tc,Gunaydin:1987hb}. There exists another singleton supermultiplet in which the roles of two spinor representations of $SO(N)$ are interchanged. For odd $N$ both the conformal scalars and conformal space-time spinors transform in the same  spinor representation of $SO(N)$.

\section{The quasiconformal realization of the minimal unitary representation of $SO(4,2)$ in compact three-grading }
\label{app-c3grading}
In this appendix we provide the formulas for the quasiconformal realization of generators of $SO(4,2)$ in compact 3-grading following \cite{Fernando:2009fq}. Consider the compact three graded decomposition of the Lie algebra of $SU(2,2)$ determined by the conformal Hamiltonian
\be
\mathfrak{so}(4,2) = \mathfrak{C}^- \oplus \mathfrak{C}^0  \oplus \mathfrak{C}^+ \nn \ee
where $\mathfrak{C}^0 =\mathfrak{so}(4)\oplus
\mathfrak{so}(2)$.
Folllowing \cite{Fernando:2009fq} we shall label the generators in $\mathfrak{C}^\pm$ and $\mathfrak{C}^0$ subspaces as follows:
\bea
 ( B_1,  B_2 , B_3 , B_4 ) \in \mathfrak{C}^-\\
 L_{\pm,0} \oplus H  \oplus R_{\pm,0} \in \mathfrak{C}^0\\
( B^1,  B^2 , B^3 , B^4 ) \in \mathfrak{C}^+
\eea
where $ L_{\pm,0}  $ and $  R_{\pm,0}$ denote the generators of $SU(2)_L\times SU(2)_R$ and $H$ is the $U(1)$ generator.

The generators  of $\mathfrak{so}(4,2)$ in the compact  3-grading take on very simple forms when expressed in terms of the singular oscillators introduced in section \ref{sect4dqc}:
\be
H = \half \lb A^\dagger_{\ml+1}A_{\ml+1} + \ml +\frac{1}{2} (N_d + N_g ) + \frac{5}{2} \rb
\ee

\be
L_+ = -\frac{i}{2} A_\ml d^\dagger, \quad L_- = \frac{i}{2} d A_\ml^\dagger , \quad L_3 = N_d - \half \lb H  - 1 \rb
\ee

\be
R_+ = \frac{i}{2} g ^\dagger A_{-\ml} , \quad R_- = -\frac{i}{2} A_{-\ml}^\dagger g , \quad R_3 = N_g - \half \lb H  + 1 \rb
\ee

\be
B_1 = -i A_\ml A_{-\ml}, \quad B_2 = -i \sqrt{2} d A_{-\ml}, \quad B_3 = -i \sqrt{2}  A_{\ml} g, \quad B_4 = -2i g d
\ee

\be
B^1 = i A^\dagger_{-\ml} A^\dagger_{\ml}, \quad B^2 = i \sqrt{2} A^\dagger_{-\ml} d^\dagger, \quad B^3 = i \sqrt{2} g^\dagger  A^\dagger_{\ml}, \quad B^4 = 2i g^\dagger d^\dagger
\ee


\section{Commutation relations of deformed twistorial oscillators}
\label{app-yztwist}

\begin{align}
\comm{Y^1}{\widetilde{Y}^1} &= \frac{1}{2x}(Y^1-\widetilde{Y}^1), & \comm{Y^2}{\widetilde{Y}^2} &= -\frac{1}{2x}(Y^2-\widetilde{Y}^2) \\
\comm{Y^1}{Y^2} &= \frac{1}{2x}(Y^1+Y^2), & \comm{\widetilde{Y}^1}{\widetilde{Y}^2} &= \frac{1}{2x}(\widetilde{Y}^1+\widetilde{Y}^2) \\
\comm{Y^1}{\widetilde{Y}^2} &= -\frac{1}{2x}(Y^1+\widetilde{Y}^2), & \comm{\widetilde{Y}^1}{Y^2} &= -\frac{1}{2x}(\widetilde{Y}^1+Y^2)
\end{align}

\begin{align}
\comm{Z_1}{\widetilde{Z}_1} &= -\frac{1}{2x}(Z_1+\widetilde{Z}_1), & \comm{Z_2}{\widetilde{Z}_2}& = -\frac{1}{2x}(Z_2+\widetilde{Z}_2) \\
\comm{Z_1}{Z_2} &= -\frac{1}{2x}(Z_1-Z_2), & \comm{\widetilde{Z}_1}{\widetilde{Z}_2} &= \frac{1}{2x}(\widetilde{Z}_1 - \widetilde{Z}_2) \\
\comm{Z_1}{\widetilde{Z}_2} &= -\frac{1}{2x}(Z_1+\widetilde{Z}_2), & \comm{\widetilde{Z}_1}{Z_2} &= \frac{1}{2x}(\widetilde{Z}_1 + Z_2)
\end{align}

\begin{align}
\comm{Y^1}{Z_1} &=2 +\frac{1}{2x}(Y^1-Z_1), & \comm{\widetilde{Y}^1}{\widetilde{Z}_1} &=2- \frac{1}{2x}(\widetilde{Y}^1 + \widetilde{Z}_1) \\
\comm{Y^1}{Z_2} &= \frac{1}{2x}(Y^1-Z_2), & \comm{\widetilde{Y}^1}{\widetilde{Z}_2} &= -\frac{1}{2x}(\widetilde{Y}^1+\widetilde{Z}_2) \\
\comm{Y^1}{\widetilde{Z}_1} &= \frac{1}{2x}(Y^1+\widetilde{Z}_1), & \comm{\widetilde{Y}^1}{Z_1} &= -\frac{1}{2x}(\widetilde{Y}^1-Z_1) \\
\comm{Y^1}{\widetilde{Z}_2} &= \frac{1}{2x}(Y^1+\widetilde{Z}_2), & \comm{\widetilde{Y}^1}{Z_2} &= -\frac{1}{2x}(\widetilde{Y}^1-Z_2)
\end{align}

\begin{align}
\comm{Y^2}{Z_1} &= \frac{1}{2x}(Y^2+Z_1), & \comm{\widetilde{Y}^2}{\widetilde{Z}_1} &=-\frac{1}{2x}(\widetilde{Y}^2 - \widetilde{Z}_1) \\
\comm{Y^2}{Z_2} &= 2+\frac{1}{2x}(Y^2+Z_2), & \comm{\widetilde{Y}^2}{\widetilde{Z}_2} &= 2-\frac{1}{2x}(\widetilde{Y}^2-\widetilde{Z}_2) \\
\comm{Y^2}{\widetilde{Z}_1} &=  \frac{1}{2x}(Y^2-\widetilde{Z}_1), & \comm{\widetilde{Y}^2}{Z_1} &= -\frac{1}{2x}(\widetilde{Y}^2+Z_1) \\
\comm{Y^2}{\widetilde{Z}_2} &= \frac{1}{2x}(Y^2-\widetilde{Z}_2), & \comm{\widetilde{Y}^2}{Z_2} &= -\frac{1}{2x}(\widetilde{Y}^2+Z_2)
\end{align}
\end{appendices}
\providecommand{\href}[2]{#2}\begingroup\raggedright\endgroup

\end{document}